\RequirePackage{fix-cm}
\documentclass[twocolumn]{svjour3}          % twocolumn
\smartqed  % flush right qed marks, e.g. at end of proof
\usepackage{graphicx}
\usepackage{mathptmx}
\usepackage{cite}
\usepackage{hyperref}
\hypersetup{
    colorlinks = true,
    citecolor = blue,
    linkcolor = red,
}
\usepackage{filecontents,lipsum}
\usepackage{subcaption}
\usepackage[skip=2pt,font=scriptsize]{caption}
\usepackage[skip=2pt,font=scriptsize]{subcaption}
\usepackage{amsmath,amssymb,amsfonts}
\usepackage{textcomp}
\usepackage{color}
\usepackage[dvipsnames]{xcolor}
\begin{document}

\title{Study the Bifurcations of a 2DoF Mechanical Impacting System
%\thanks{This work was supported by DST--INSPIRE Fellowship (Ref. No. IF150667) and J C Bose Fellowship (Ref. no. SB/S2/JCB-023/2015), Government of India.}
}
%\subtitle{}

\titlerunning{Study the dynamics of a 2D Mechanical Impacting System}        % if too long for running head

\author{Soumyajit Seth \and Grzegorz Kudra \and Grzegorz Wasilewski \and Jan Awrejcewicz}

\authorrunning{Seth et al.} % if too long for running head

\institute{S. Seth*\and G. Kudra \and G. Wasilewski \and J. Awrejcewicz \at
              Department of Automation, Biomechanics and Mechatronics, Lodz University of Technology, Poland.\\  *Corresponding author \\
            \email{soumyajit.seth@p.lodz.pl}
            \and
            G. Kudra \at
            grzegorz.kudra@p.lodz.pl
            \and
            G. Wasilewski \at
            gwasilew@p.lodz.pl
            \and
            J. Awrejcewicz \at
            jan.awrejcewicz@p.lodz.pl
}

\date{Received: date / Accepted: date}

\maketitle

\begin{abstract}
Impacting mechanical systems with suitable parameter settings exhibit a large amplitude chaotic oscillation close to the grazing with the impacting surface. The cause behind this uncertainty is the square root singularity and the occurrence of dangerous border collision bifurcation. In the case of one degree of freedom mechanical systems, it has already been shown that this phenomenon occurs under certain conditions. This paper proposes the same uncertainty of a two-degree freedom mechanical impacting system under specific requirements. This paper shows that the phenomena earlier reported in the case of one degree of freedom mechanical systems (like narrow band chaos, finger-shaped attractor, etc.) also occur in the two degrees of freedom mechanical impacting system. We have numerically predicted the narrowband chaos ensues under specific parameter settings. We have also shown that the narrowband chaos can be avoided under some parameter settings. At last, we demonstrate the numerical predictions experimentally by constructing an equivalent electronic circuit of the mechanical rig.

\keywords{Mechanical Impacting System \and Square Root Singularity \and Narrow Band Chaos \and Electronic Switching Systems.}
\end{abstract}

\section{Introduction}
\label{intro}
Various dynamical systems are observed in multiple areas of science and engineering, where impacts occur between the components of the systems. These systems exhibit a rich sort of dynamical phenomena, especially in the range of parameter values where grazing occurs. The phenomena include transitioning from one periodic attractor to another through the chaotic orbit at grazing, finger-shaped chaotic attractors at the bifurcation point in the Poincar\'e section, etc. These practical and engineering systems have been studied in detail for the last thirty years, especially on one degree of freedom mechanical impacting system under different configurations \cite{budd1995grazing, awrejcewicz2003bifurcation, ing2006dynamics, ing2007experimental, pavlovskaia2004analytical, pavlovskaia2004two, shaw1983periodically, thota2006continuous, witkowski2019modelling}.

The bifurcation structure of impact oscillators was investigated by Feigin \cite{feigin1978structure}. Whiston \cite{whiston1987global} provided a numerical approach to study the dynamics, such as the steady-state analysis, phase-space diagrams, domains of attraction, etc., of a one-degree of freedom Vibro-mechanical impacting system with and without excitations. Nordmark \cite{nordmark1991non} showed the characteristics `Square root singularity' at the grazing condition where one of the Jacobian elements of the Poincar\'e map goes to infinity, which leads to infinite local stretching in the state space. Peterka et al. \cite{peterka1992transition}, Ivanov \cite{ivanov1993stabilization}, and Lenci et al. \cite{lenci1998procedure} showed the transition to chaos, its stabilization, and reduction of chaos at the grazing condition of a mechanical impacting system. Blazejczyk-Okolewska et al. have shown the co-existing attractors in impacting systems having dry friction under the influence of system noise \cite{blazejczyk1998co}. Dankowicz et al. \cite{dankowicz2000origin} have discussed the stability analysis and the bifurcations of a periodic orbit associated with the stick-slip oscillations. Bernardo et al. \cite{di2001normal, di2002bifurcations} presented the unified framework of local analysis of grazing and sliding bifurcations. They showed that this leads to a normal map form under some general conditions, which contains a lower order square-root singularity or a $\frac{3}{2}$--singularity. Awrejcewicz et al. \cite{awrejcewicz2002nonlinear} studied numerically the dynamics of a triple pendulum having an impact and showed that under certain conditions, periodic, quasi-periodic, and chaotic motions were detected. Ma et al.\cite{ma2006border} have studied the occurrence of large amplitude chaos of a soft impacting system and showed that in the case of a discrete map, during the bifurcation, the determinant of the Jacobian matrix remains invariant, and the trace shows a singularity at the grazing point. They have also shown how the character of a soft impacting system's zero time discontinuity map changes over a range of parameters as the system is driven from a non-impacting orbit to an impacting orbit \cite{ma2008nature}. Ing et al. \cite{ing2006dynamics, ing2007experimental} showed the bifurcation scenarios close to grazing experimentally for a nearly symmetrical piecewise linear mechanical impacting oscillator. They also have experimentally studied the bifurcations of an impact oscillator with a one-sided elastic constraint. Banerjee et al. \cite {banerjee2009invisible} have discovered a narrow band of chaos close to the grazing condition for a simple soft impact oscillator experimentally for a range of system parameters. Also, numerical stability analysis shows that this abrupt onset of chaos is caused by a dangerous bifurcation where two unstable period-$3$ orbits take part at grazings. Kundu et al.\cite{kundu2012singularities} have numerically investigated the character of the normal form map in the neighborhood of a grazing orbit for four possible configurations of soft impacting systems. They have shown the conditions when there is an onset of chaos and under which this onset of chaos can be avoided for the one degree of freedom mechanical impacting systems. Suda and Banerjee \cite{suda2016does} have shown that for one degree of freedom mechanical impacting system, one can avoid the narrow band chaos not just for singular parameter values but for a range of parameter values. They have demonstrated its mechanism by computing the interplay between stable and unstable periodic orbits in the bifurcation diagram. George et al. \cite{george2016experimental} showed some typical behaviors such as finite-time transient behavior of the orbit before settling to a long-time behavior of an impacting mechanical system. Witkowski et al. \cite{witkowski2019modelling} have studied the dynamics of a mechanical one-degree-of-freedom oscillator with harmonic forcing and impacts both numerically and experimentally. They have shown the bifurcation diagram obtained experimentally where a transition of chaos occurs from a periodic orbit under the variation of a parameter. Seth and Banerjee \cite{seth2020electronic} have proposed an electronic switching circuit that can act as an analog of a one-degree of freedom mechanical impacting system. They have shown that the phenomena reported earlier through numerical simulation (like narrow-band chaos, finger-shaped attractor, etc.) also occur in the circuit. They have experimentally obtained the evolution of the chaotic attractor at grazing as the stiffness ratio varies, which is very hard to perform in mechanical rigs. They also have confirmed experimentally that the theoretical prediction of the occurrence of narrow-band chaos can be avoided for some discrete values of parameters.

So, most of the works have been completed from theoretical points of view. The phenomena have been extensively studied experimentally in one degree of freedom mechanical impacting systems. However, it is required to investigate whether the phenomena, such as narrow-band chaos and finger-shaped attractor, may occur in two-degree of freedom mechanical impacting systems or not. 

The purpose of this paper is to propose a forced two-degrees of freedom mechanical system with a compliant impact. Under some parameter settings, we have shown the onset of chaos in the bifurcation diagram when the amplitude of the externally applied periodic signal is varied. We also have shown that when there is a specific relation between the externally applied signal's frequency and the system's natural frequencies, the chaotic attractor can be avoided, as studied in one degree of freedom mechanical impacting system. We have fabricated an electronic circuit analogous to the mechanical oscillator. We have confirmed that the dynamical phenomena observed in the one degree of freedom mechanical systems are also observed in the two degrees of freedom system, both numerically and experimentally.

\section{Mechanical system under investigation}
\label{mech_sys}
\subsection{System description}
\label{sys_des}
\begin{figure}[tbh]
\centering
\includegraphics[width=3.1in]{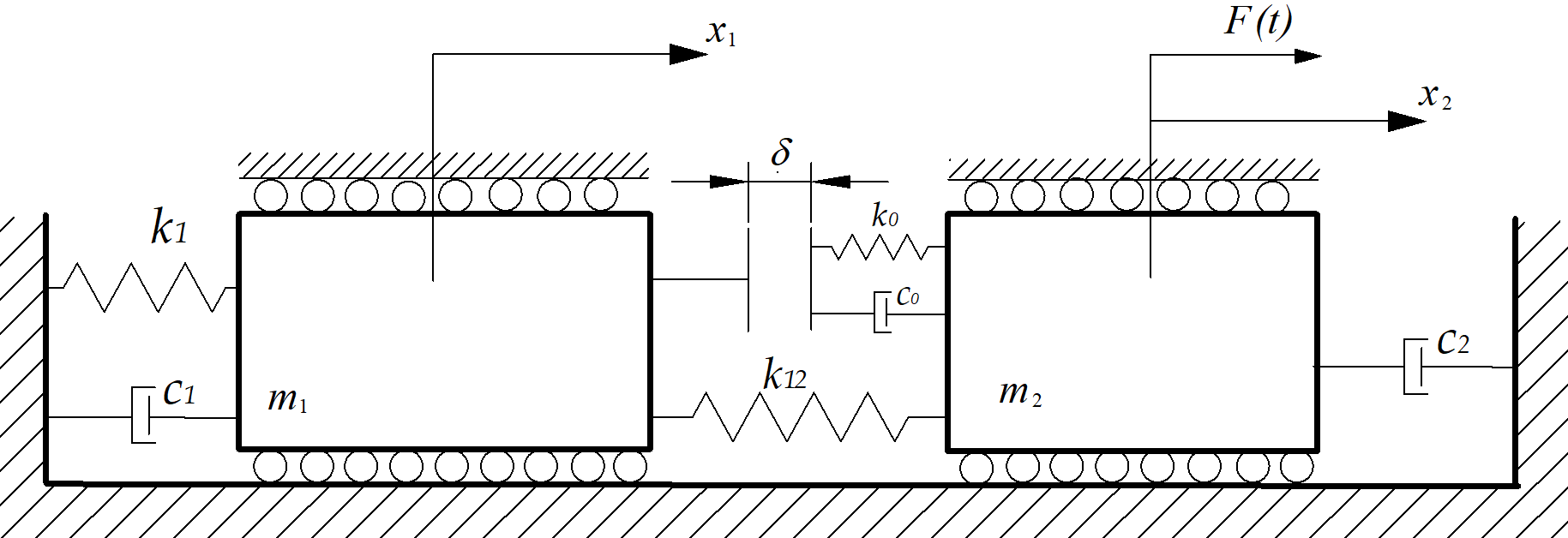}
\caption{The schematic representation of a two-degree of freedom mechanical impacting system.}
\label{mech}
\end{figure}
A schematic diagram of a two-degree of freedom mechanical system under study is depicted in Fig.~\ref{mech}. It is a forced damped oscillator with a massless compliant obstacle that the mass $m_1$ can impact. The massless complaint obstacle is attached to another mass $m_2$ with a spring having spring constant $k_o$ and a damper with the damping coefficient $c_o$. The mass $m_1$ is attached to a fixed support by a spring with spring constant $k_1$ and a damper $c_1$. The mass $m_2$ is connected to the fixed support with a damper $c_2$. The two masses $m_1$ and $m_2$ are connected with a spring having spring constant $k_{\rm 12}$. The forcing periodic function $F(t)$ is acted on the mass $m_2$. Due to the application of the external forcing, the masses $m_1$ and $m_2$ started to oscillate from their equilibrium positions. $x_1$ and $x_2$ are the amount of displacements of the two masses $m_1$ and $m_2$ from their equilibrium positions, respectively. When no external force is acting on $m_2$, the linear distance between the complaint and $m_1$ at equilibrium is $\delta$. If there is a force acting on $m_2$, two situations will come to an existing. First, there will be no impact. In this situation, the masses $m_1$ and $m_2$ will oscillate with the force $F(t)$. The second possibility is when there will be an impact between $m_1$ and the compliant wall. The grazing condition will happen when the mass $m_1$ touches the impacting surface attached with the mass $m_2$. In our work, we have neglected the friction terms related to the movements of the two masses with the hard surface.

\subsection{Mathematical equations of the model}
\label{mech_math}
Considering the dry-friction between the masses and the hard surface has negligible effect on $x_1$ and $x_2$, we can write the equations of motion of the considered system as below:

For $(x_2 - x_1) < \delta$,
\begin{equation}
\begin{split}
m_1\ddot{x}_1 &= - c_1\dot{x}_1 - k_1x_1 - k_{\rm 12}(x_1-x_2) \\
m_2\ddot{x}_2 & = - c_2\dot{x}_2 +  k_{\rm 12}(x_1-x_2) + F(t)
\end{split}
\label{eq1}
\end{equation}

and, for $(x_2- x_1) \geq \delta$
\begin{equation}
\begin{split}
m_1\ddot{x}_1 = &- c_1\dot{x}_1 - k_1 x_1 -  k_{\rm 12}(x_1-x_2) - c_0 (\dot{x}_1 - \dot{x}_2) \\& - k_0 (x_1 - x_2 - \delta) \\
m_2\ddot{x}_2  = &- c_2\dot{x}_2 +  k_{\rm 12}(x_1-x_2) + c_0 (\dot{x}_1 - \dot{x}_2) \\& + k_0 (x_1 - x_2 - \delta) + F(t)
\end{split}
\label{eq2}
\end{equation}
where, $F(t) = F_o \sin (\Omega t)$ is the external excitation to the mass $m_2$.

\subsection{Non-dimensional form of the system}
\label{non_form}
In order to study the dynamics further of the considered system, we want to transform the dimensional equations~(\ref{eq1}) and (\ref{eq2}) to the non-dimensional one. We have introduced non-dimensional time as $\tau = \omega_n t$, where, $\omega_{\rm n} = \sqrt{\frac{K_{\rm 1}}{m_{\rm 1}}}$. So, in the case of non-dimensional equations, the derivative has been chosen with respect to the non-dimensional time, $\tau$. Now, we introduce the expression of the non-dimensional forcing frequency as, $\omega = \frac{\Omega}{\omega_{\rm n}}$. The non-dimensional state-variables are defined as $y_{\rm i} = \frac{x_{\rm i}}{\delta}$; $i = 1,2$.

Considering, $\frac{dy}{d\tau} = y^{\prime}$, and, $\frac{d^2y}{d\tau^2} = y^{\prime\prime}$, we can write, $\dot{x}_{\rm i} = \frac{dx_{\rm i}}{dt}$ = $\frac{dx_{\rm i}}{d\tau}\cdot  \frac{d\tau}{dt}$ = $\omega_{\rm n} \delta y^{\prime}_{\rm i}$; similarly, $\ddot{x}_i$ = $\omega_{\rm n}^2 \delta y^{\prime \prime}_i$.

So, from the equation~(\ref{eq1}), for the condition, $(y_2 - y_1) < 1$
\begin{equation}
\begin{split}
m_1 \omega_{\rm n}^2 \delta y^{\prime \prime}_1 = & - c_1 \omega_{\rm n} \delta y^{\prime}_1 - k_1 \delta y_1 - k_{\rm 12} \delta (y_1 - y_2)\\
m_2\omega_{\rm n}^2 \delta y^{\prime \prime}_2 = & - c_2\omega_{\rm n} \delta y^{\prime}_2 +  k_{\rm 12} \delta (y_1-y_2) + F_0 \sin(\omega \tau)
\end{split}
\label{eq3}
\end{equation}
Similarly, from the equation~(\ref{eq2}), with have a condition in non-dimensional case $(y_2 - y_1) \geq 1$,
\begin{equation}
\begin{split}
m_1 \omega_{\rm n}^2 \delta y^{\prime \prime}_1 = &- c_1\omega_{\rm n} \delta y^{\prime}_1 - k_1 \delta y_1 -  k_{\rm 12} \delta (y_1 - y_2) \\& - c_0  \omega_{\rm n} \delta (y^{\prime}_1 - y^{\prime}_2) - k_0 \delta (y_1 - y_2 - 1) \\
m_2\omega_{\rm n}^2 \delta y^{\prime \prime}_2  = &- c_2 \omega_{\rm n} \delta y_2^\prime +  k_{\rm 12} \delta (y_1 - y_2) + c_0 \omega_{\rm n} \delta (\dot{y}_1 - \dot{y}_2) \\& + k_0 \delta (y_1 - y_2 - 1) + F_0 \sin(\omega \tau)
\end{split}
\label{eq4}
\end{equation}

Dividing the upper and the lower equations of (\ref{eq3}) and (\ref{eq4}) by $m_1 \omega_{\rm n}^2 \delta$ and $m_2 \omega_{\rm n}^2 \delta$, respectively, we get,

For, $(y_2 - y_1) < 1$
\begin{equation*}
\begin{split}
y^{\prime \prime}_1 = & - \frac{c_1}{m_1 \omega_{\rm n}} y^{\prime}_1 - \frac{k_1}{m_1 \omega_{\rm n}^2} y_1 - \frac{k_{\rm 12}}{m_1 \omega_{\rm n}^2} (y_1 - y_2)\\
y^{\prime \prime}_2 = & - \frac{c_2}{m_2\omega_{\rm n}} y^{\prime}_2 +  \frac{k_{\rm 12}}{m_2\omega_{\rm n}^2} (y_1-y_2) + \frac{F_0}{m_2\omega_{\rm n}^2 \delta} \sin(\omega \tau)
\end{split}
\label{eq5}
\end{equation*}
and, for, $(y_2 - y_1) \geq 1$,
\begin{equation*}
\begin{split}
y^{\prime \prime}_1 = & - \frac{c_1}{m_1 \omega_{\rm n}} y^{\prime}_1 - \frac{k_1}{m_1 \omega_{\rm n}^2} y_1 -  \frac{k_{\rm 12}}{m_1 \omega_{\rm n}^2} (y_1 - y_2) \\& - \frac{c_0}{m_1 \omega_{\rm n}} (y^{\prime}_1 - y^{\prime}_2) - \frac{k_0}{m_1 \omega_{\rm n}^2 } (y_1 - y_2 - 1) \\
y^{\prime \prime}_2  = &- \frac{c_2}{m_2\omega_{\rm n}^2} y^{\prime}_2 +  \frac{k_{\rm 12}}{m_2\omega_{\rm n}^2}(y_1 - y_2) + \frac{c_0}{m_2\omega_{\rm n}} (\dot{y}_1 - \dot{y}_2) \\& + \frac{k_0}{m_2\omega_{\rm n}^2} (y_1 - y_2 - \delta) + \frac{F_0}{m_2\omega_{\rm n}^2 \delta} \sin(\omega \tau)
\end{split}
\label{eq6}
\end{equation*}

From the above equations using the expression $\omega_n = \sqrt{\frac{k_1}{m_1}}$, we can write,

For, $(y_2 - y_1) < 1$
\begin{equation}
\begin{split}
y^{\prime \prime}_1 = & - 2\xi_1 y^{\prime}_1 - y_1 - \beta (y_1 - y_2)\\
y^{\prime \prime}_2 = & - 2 \xi_2 y^{\prime}_2 +  \frac{\beta}{\mu} (y_1-y_2) + f_0 \sin(\omega \tau)
\end{split}
\label{eq7}
\end{equation}
and, for, $(y_2 - y_1) \geq 1$,
\begin{equation}
\begin{split}
y^{\prime \prime}_1 = & - 2 \xi_1 y^{\prime}_1 - y_1 - \beta (y_1 - y_2) - 2 \xi_0 (y^{\prime}_1 - y^{\prime}_2) \\& - \beta_0 (y_1 - y_2 - 1) \\
y^{\prime \prime}_2  = &- 2\xi_2 y^{\prime}_2 +  \frac{\beta}{\mu}(y_1 - y_2) + \frac{2 \xi_0}{\mu} (y^{\prime}_1 - y^{\prime}_2) \\& + \frac{\beta_0}{\mu} (y_1 - y_2 - 1) + f_0 \sin(\omega \tau)
\end{split}
\label{eq8}
\end{equation}
 where, $\xi_1 = \frac{c_1}{2 m_1 \omega_n}$, $\xi_2 = \frac{c_2}{2 m_2 \omega_n}$, $\xi_0 = \frac{c_0}{2 m_1 \omega_n}$, $\beta = \frac{k_{\rm 12}}{k_1}$, $\beta_0 = \frac{k_0}{k_1}$, $\mu = \frac{m_2}{m_1}$, and $f_0 = \frac{F_0}{m_2 {\omega_n}^2 \delta}$.
   
\subsection{Expressions of the natural frequencies}
\label{nf}
As our considered system is two-dimensional, it has two normal modes of vibration corresponding to two natural frequencies. From the equations~(\ref{eq7}), we can calculate the natural frequencies ($\omega_{\rm +}$ and $\omega_{\rm -}$) of the system. Considering the free vibration analysis of the system, we can disregarded $\xi_1$ = $\xi_2$ = $f_0$ = $0$, and the equations of motion are reduced to
\begin{equation*}
\begin{split}
y^{\prime \prime}_1 = & - y_1 - \beta (y_1 - y_2)\\
y^{\prime \prime}_2 = & +  \frac{\beta}{\mu} (y_1-y_2)
\end{split}
\label{eq_nf}
\end{equation*}

So, the natural frequencies are expressed as,
\begin{equation}
\omega_{\rm \pm} = \pm \frac{1}{\sqrt{2}}\Bigg[\Big\{\beta \Big(1 + \frac{1}{\mu}\Big) + 1\Big\} \pm \sqrt{\Big\{\beta \Big(1 + \frac{1}{\mu} \Big) +1\Big\}^{2} - \frac{4\beta}{\mu}}\Bigg]^\frac{1}{2}
\end{equation}
We can calculate the natural frequencies from the above equations by choosing the parameter values of $\beta$ and $\mu$.

In the case of a one-degree of freedom mechanical impacting system, we have observed the onset of large amplitude chaotic oscillation at the bifurcation point when the forcing frequency is not an integer multiple of twice the natural frequency of the system. Also, this onset of chaos can be avoided when the external excitation frequency has a definite relation with the natural frequency of the system \cite{kundu2012singularities, banerjee2009invisible}. Likewise, in our work, we shall observe the occurrence of the large amplitude chaotic oscillation at the point of bifurcation when the frequency of the externally applied signal is not related to an integer with twice the average value of the two natural frequencies. Also, like previous observations, we shall find whether this large amplitude chaos can be eliminated when the relation is close to the integer one. The average value of the natural frequencies of the system is:
\begin{equation}
\omega_{\rm avg} = \frac{(\omega_{\rm +} + \omega_{\rm -})}{2}
\label{omega_average}
\end{equation}

 \section{Numerical results from nondimensional equations}
 \label{numerical}
To show the dynamics of the considered mechanical system~(\ref{mech}), we have chosen the non-dimensional forms of the mathematical equations of the system as written in Eq.~(\ref{eq7}) and Eq.~(\ref{eq8}). We have selected $f_0$ as a varying non-dimensional parameter, and the remaining parameters have been kept fixed. If we increase $f_0$ from low to high values, we can obtain different dynamics, such as various orbits of the system. In our work, we have considered the two conditions, $\frac{\omega}{\omega_{\rm avg}} = \frac{2}{m}$, (i) $m$ is not an integer, i.e., non-integer and (ii) $m$ is an integer.

\subsection{$m = 2.5106$, i.e., the non-integer condition}
\label{m_ni}
The non-dimensional parameters are chosen as  $\xi_1 = 0.014$, $\xi_2 = 0.018$, $\xi_0 = 6.6296 \times 10^{-4}$, $\beta = \frac{1}{10.4}$, $\beta_0 = 25.02$, and $\mu = \frac{1}{9.8}$. $f_0$ is considered to be varied. Using those parameter values, $\omega_{\rm avg}$ is $0.9974$. For the simplicity of the calculation, we have approximated the $\omega_{\rm avg}$ value to be unity. So, $\omega = \frac{2}{m}$.

\begin{figure}[tbh]
  \centering
  \begin{subfigure}[b]{\linewidth}
  \includegraphics[width=\linewidth]{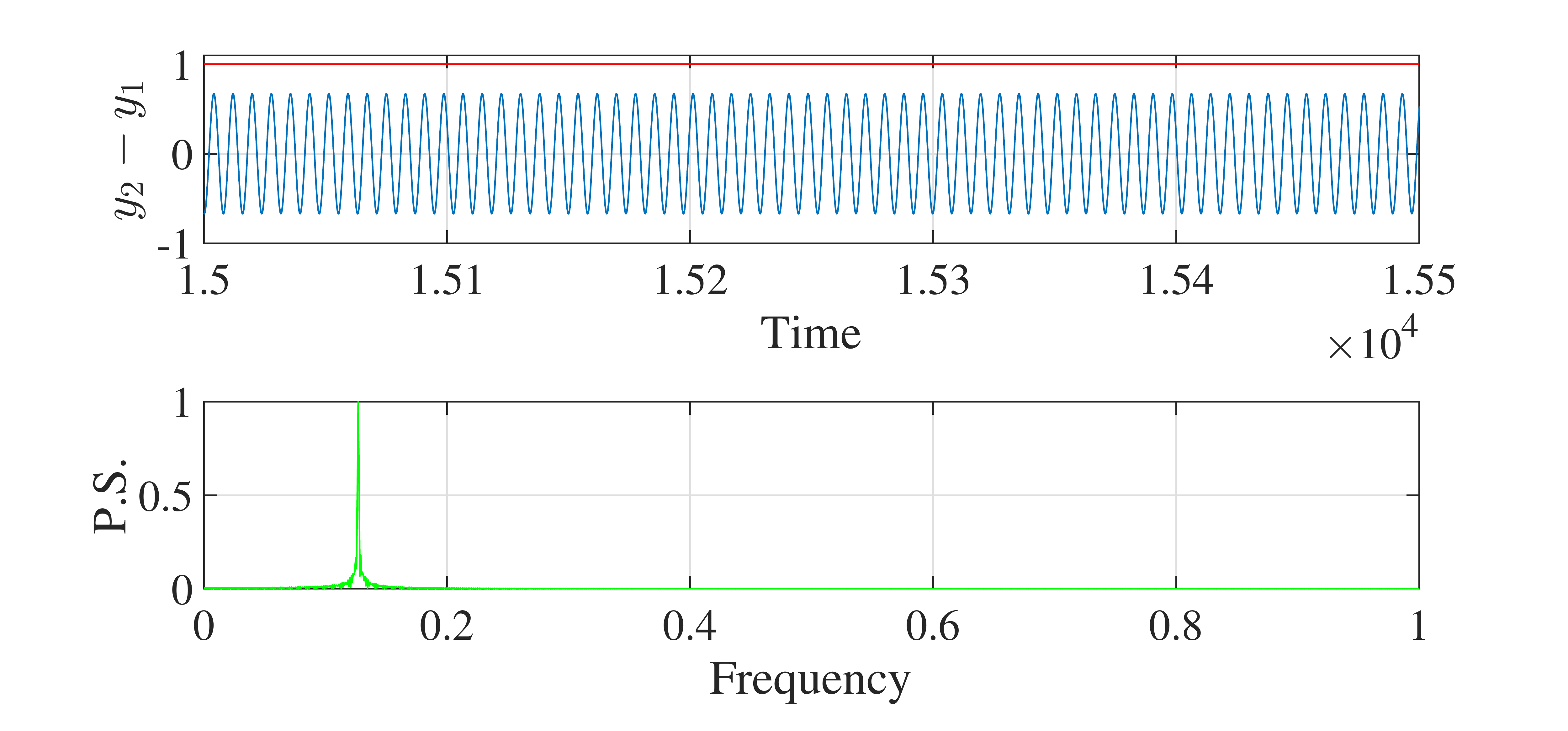}
  \caption{}
  \end{subfigure}  
  \begin{subfigure}[b]{\linewidth}
  \includegraphics[width=\linewidth]{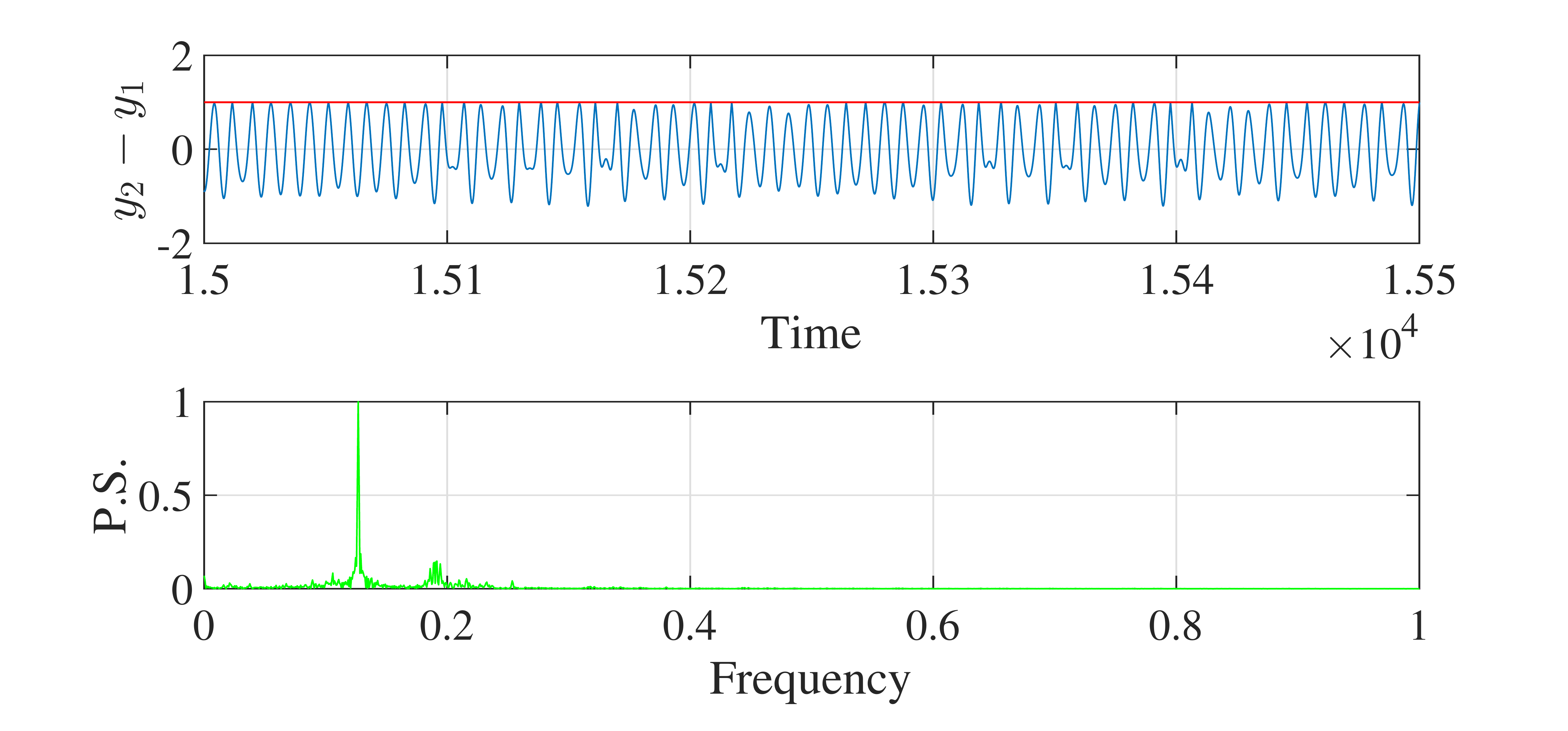}
  \caption{}
  \end{subfigure}  
  \begin{subfigure}[b]{\linewidth}
    \includegraphics[width=\linewidth]{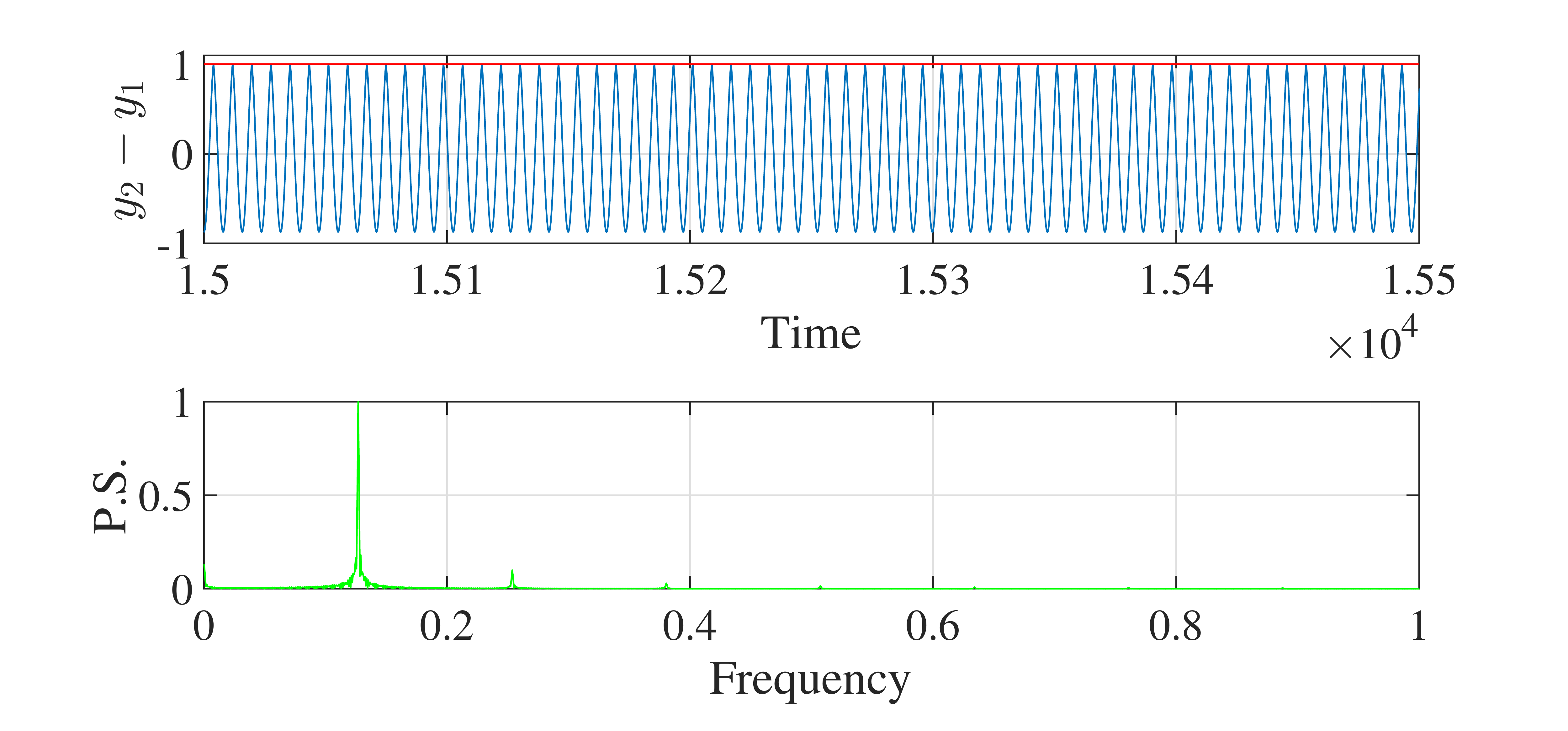}
    \caption{}
  \end{subfigure}  
  \caption{Time-series waveforms and the corresponding frequency spectra for different values of $f_0$. (a) Before impact for $f_0 = 0.1$, Period-$1$ orbit, (b) At grazing condition for $f_0 = 0.17$, Chaotic orbit, (c) After bifurcation for $f_0 = 0.22$, period-$1$ orbit. In each subfigure, the upper trace is the time-series waveforms, The $x$-axis is the non-dimensional time, and the $y$-axis is the non-dimensional displacement ($y_2 - y_1$) (in blue color) compared with a constant value, $1$ (in red color). The lower trace is the frequency spectra, where the $x$-axis is the nondimensional frequency and $y$-axis is the Power spectrum. The initial condition is chosen at $(-0.5,-0.1,-0.01,1)$. (Color online.)}
  \label{ts_nd}
\end{figure}
Fig.~\ref{ts_nd} shows different time-series waveforms for different values of $f_0$. Fig.~\ref{ts_nd}(a) shows a period-$1$ waveform and corresponding its frequency spectra. The single peak at the externally applied non-dimensional frequency, $0.1468$, confirms that the orbit is periodic. Note that, the state-variable $(y_2 - y_1)$ does not touch $1$. This is the condition before border collision. During that moment, the two masses oscillate freely. When $f_0$  is increased, the grazing will occur between the mass $m_1$ and the wall attached with the mass $m_2$. Fig.~\ref{ts_nd}(b) shows the existence of the chaotic orbit where $(y_2 - y_1)$ touches $1$, i.e. at the bifurcation point. At that point, the waveform becomes irregular. The power spectrum is distributed over a large range in the frequency axis, although there is a peak at $0.1468$. This peak is due to the applied signal frequency. As $f_0$ is increased more, a period-$1$ orbit emerges. This is the condition after the bifurcation. Fig.~\ref{ts_nd}(c) shows the time-series waveform and the corresponding frequency spectra of the period-$1$ orbit.

\begin{figure}[tbh]
  \centering
  \begin{subfigure}[b]{\linewidth}
  \includegraphics[width=\linewidth]{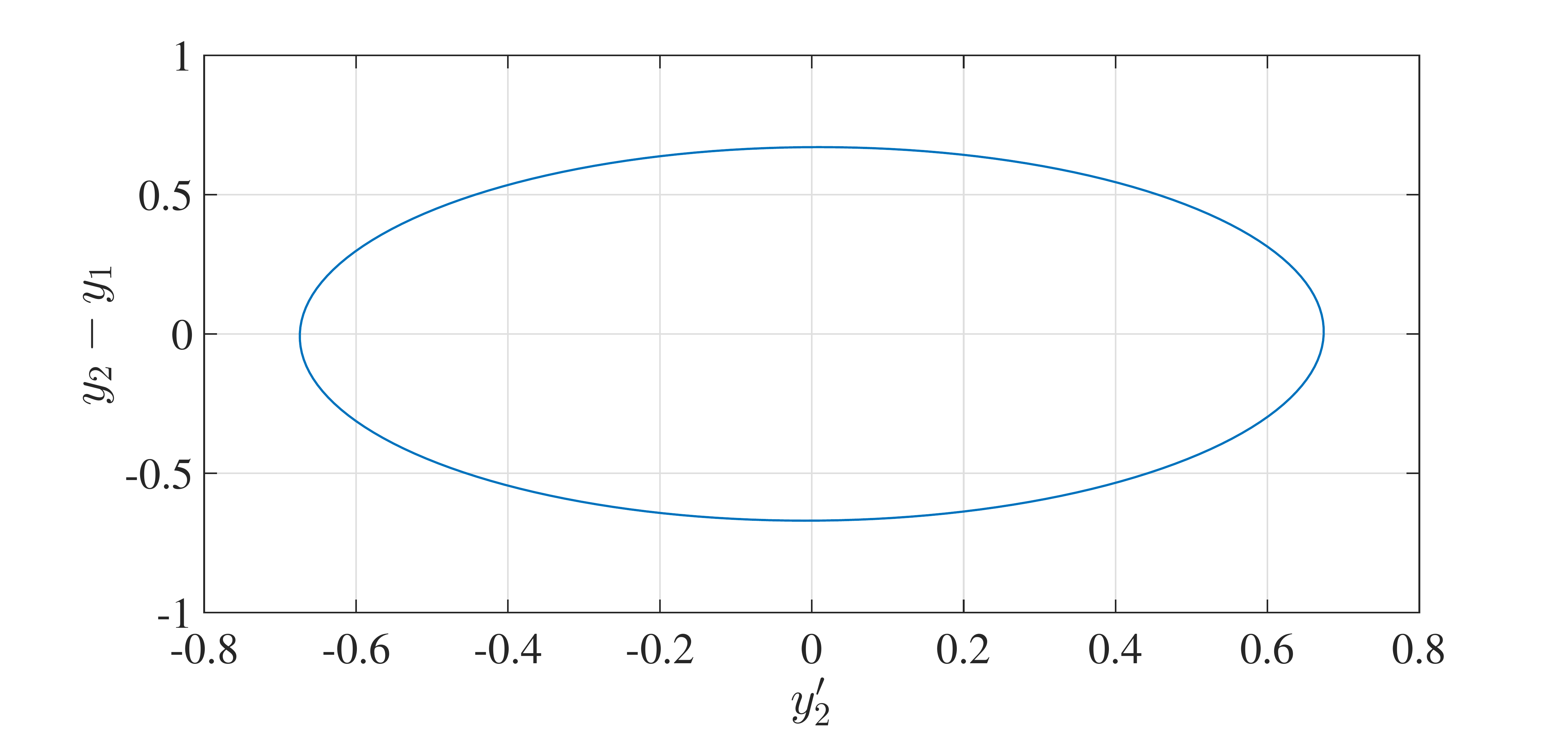}
  \caption{}
  \end{subfigure}  
  \begin{subfigure}[b]{\linewidth}
  \includegraphics[width=\linewidth]{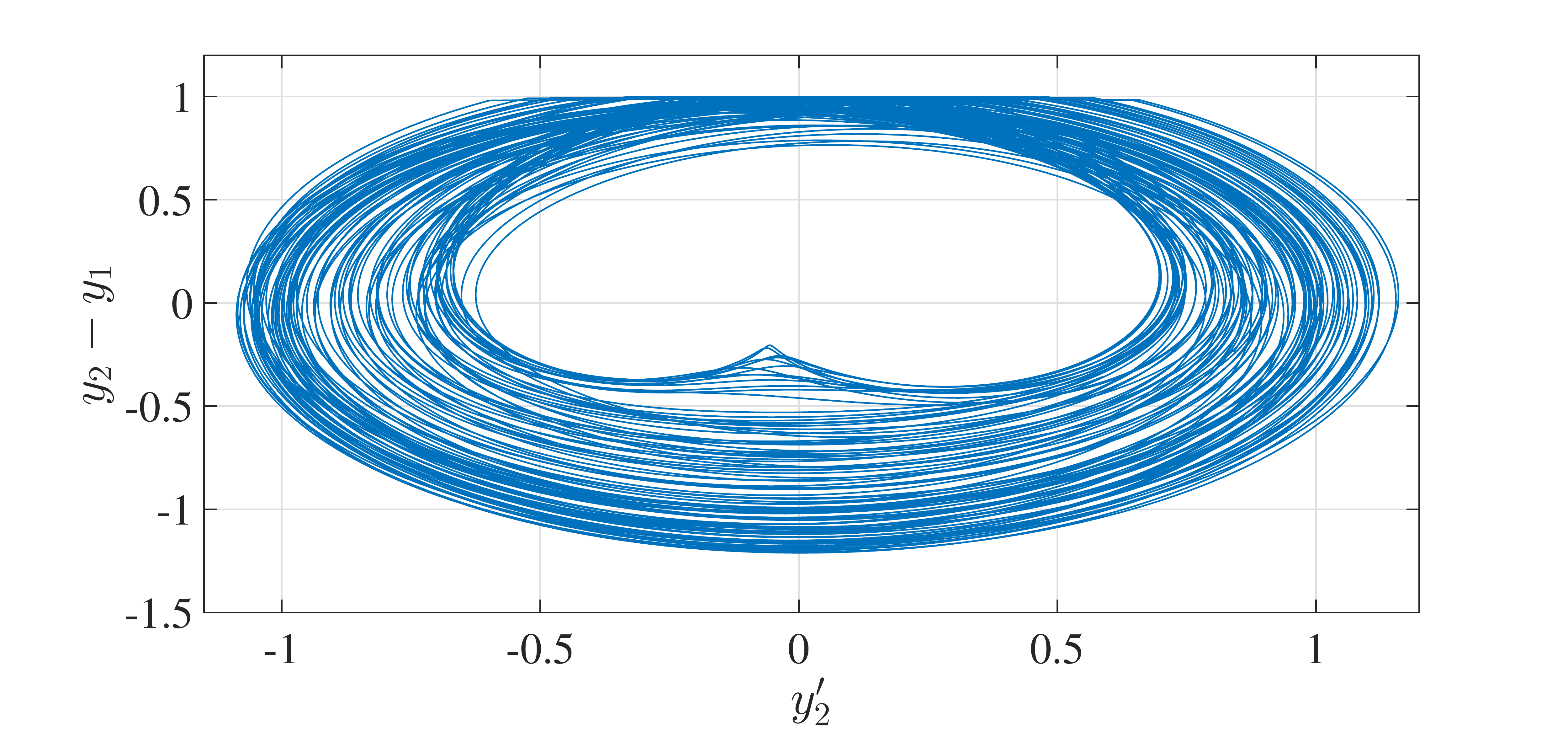}
  \caption{}
  \end{subfigure}  
  \begin{subfigure}[b]{\linewidth}
    \includegraphics[width=\linewidth]{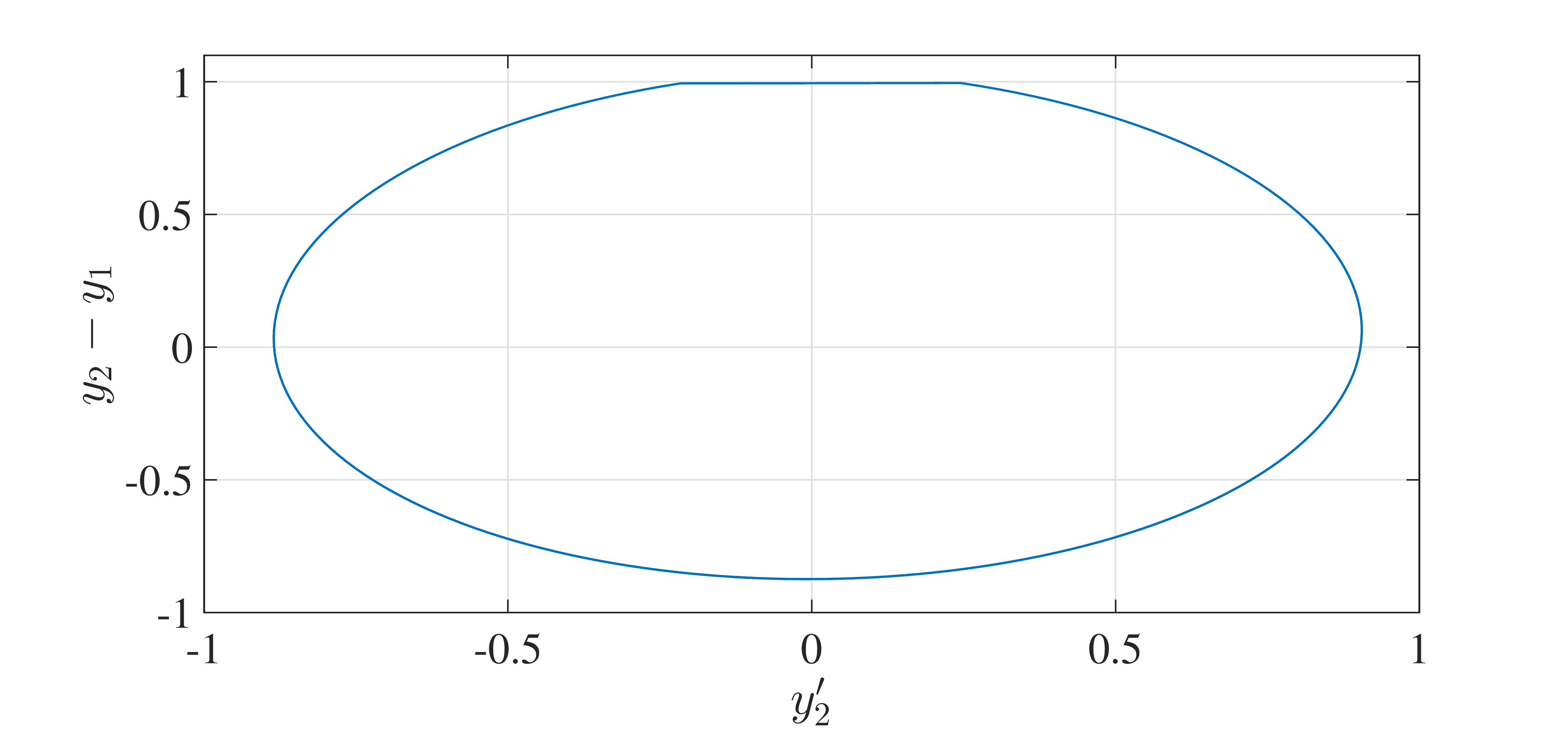}
    \caption{}
  \end{subfigure}  
  \caption{Phase-Space trajectories for different values of $f_0$. (a) Before impact for $f_0 = 0.1$, Period-$1$ orbit, (b) At grazing condition for $f_0 = 0.17$, Chaotic orbit, (c) After bifurcation for $f_0 = 0.22$, Period-$1$ orbit. The $x$-axis is the non-dimensional velocity $y^{\prime}_2$ and the $y$-axis is the non-dimensional displacement $(y_2 - y_1)$. The initial condition is chosen at $(-0.5,-0.1,-0.01,1)$. (Color online.)}
  \label{ps_nd}
\end{figure}
Fig.~\ref{ps_nd} show different phase-space trajectories for different values of $f_0$. We have chosen the same non-dimensional parameter values as previously. Fig.~\ref{ps_nd}(a) shows a period-$1$ orbit for $f_0 = 0.1$. The single loop in the phase-space makes the orbit period-$1$. This is the condition before impact. The grazing condition will happen when $f_0$  is increased more in the forward direction. For $f_0 = 0.17$, the orbit becomes chaotic. The fickle nature of the orbit in the phase-space confirms its chaotic nature. The chaotic waveform of the state-variable remains for a range of $f_0$. Fig.~\ref{ps_nd}(c) shows the phase-space diagram after the bifurcation. There is one loop in the phase-space, which makes the orbit a period of $1$.

\begin{figure}[tbh]
\centering
\includegraphics[width = \linewidth]{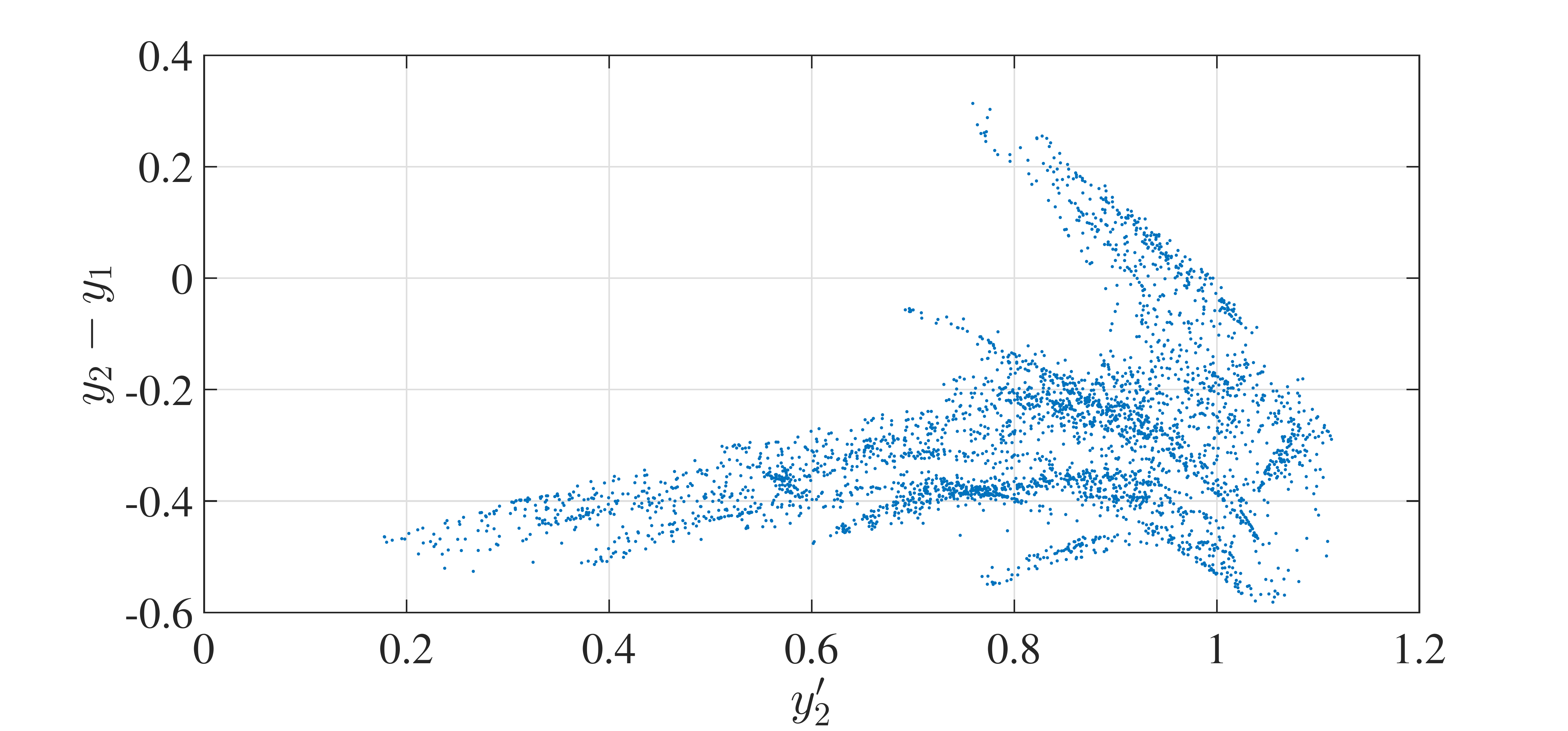}
\caption{Poincar\'e section of the chaotic attractor at grazing ($f_0 = 0.17$) for $m = 2.5106$ showing the large amplitude chaotic oscillation at the bifurcation. The $x$-axis is the non-dimensional velocity $y^{\prime}_2$ and the $y$-axis are the non-dimensional displacement $(y_2 - y_1)$. (Color Online.)}
\label{nd_ni_poincare}
\end{figure}
Fig.~\ref{nd_ni_poincare} shows discrete-time phase-space attractor at the grazing condition. The phase space is discretized with the synchronism of the period of the input sine wave, which means we shall observe the state variable at every $T$ instant of time, where $T = \frac{\omega}{2\pi}$. The chaotic attractor at the bifurcation point has a typical finger-shaped structure as shown in one degree of freedom mechanical impacting system \cite{banerjee2009invisible}. It confirms that the proposed two degrees of freedom mechanical impacting system under the considered parameter values shows the same phenomena as we have seen in a soft impacting one-degree of freedom system. The fuzzy dots in the finger-shaped attractor are coming because the impacting wall is attached with the mass $m_2$, which is moving back and forth during grazing.

\begin{figure}[tbh]
\centering
\begin{subfigure}[b]{\linewidth}
\includegraphics[width = \linewidth]{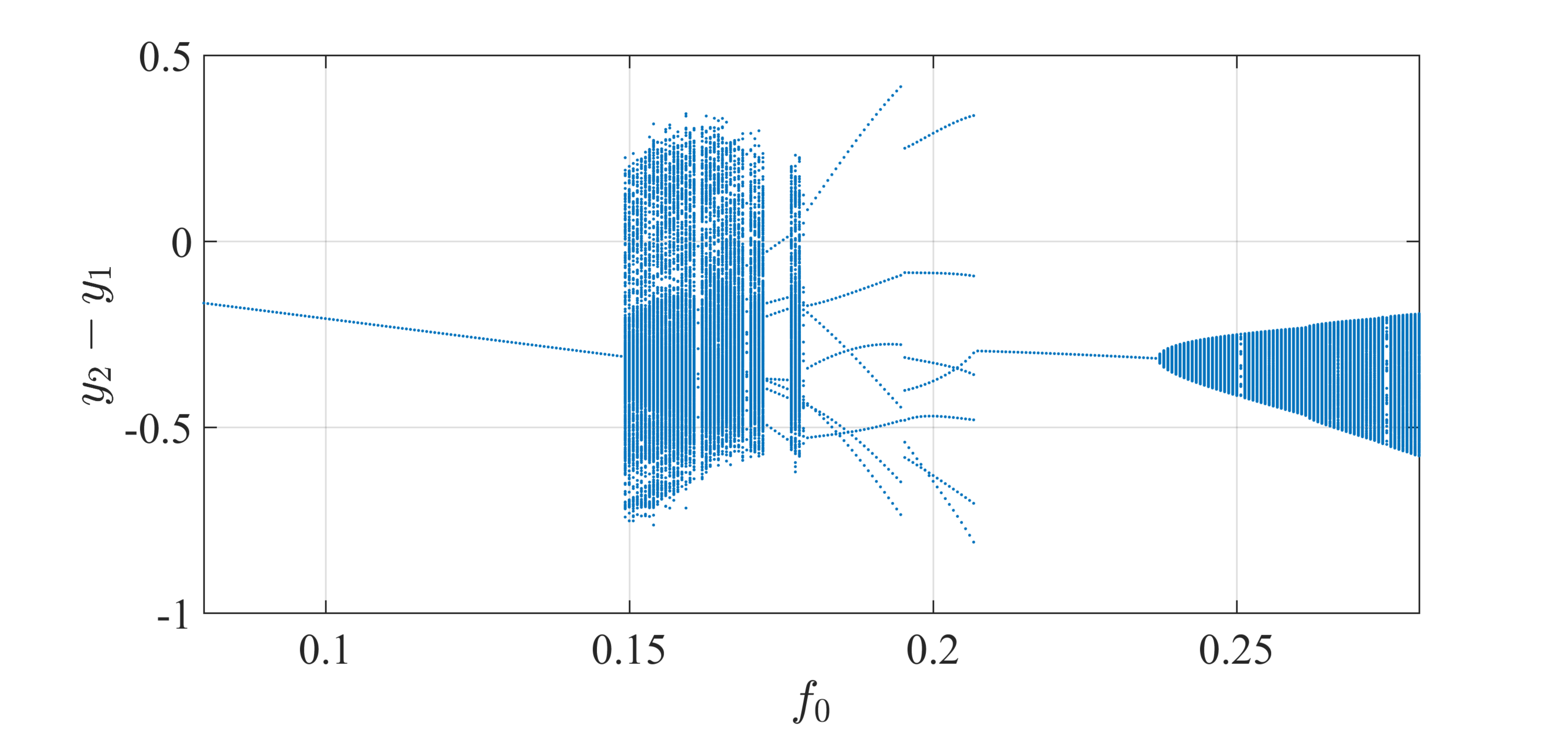}
\caption{}
\end{subfigure}
\begin{subfigure}[b]{\linewidth}
\includegraphics[width = \linewidth]{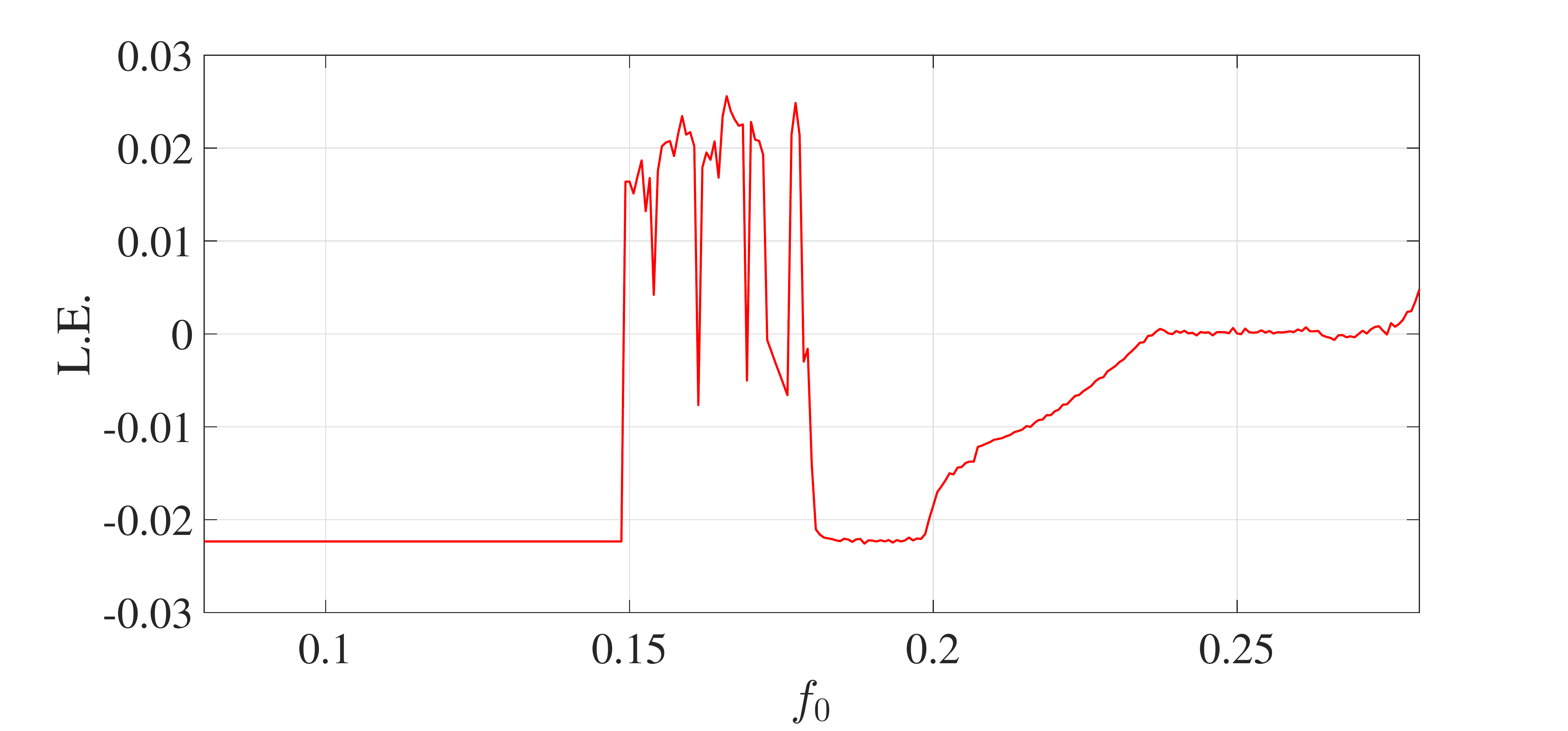}
\caption{}
\end{subfigure}
\caption{(a) Bifurcation diagram and (b) the maximal Lyapunov Exponent plot in non-dimensional parameter space of the system~(\ref{mech}) for $m = 2.5106$ showing the large amplitude chaotic oscillation in the bifurcation diagram around the bifurcation point when the bifurcation parameter $f_0$ is varied from $0.08$ to $0.28$. (Color online.)}
\label{nd_ni_bif}
\end{figure}
Fig.~\ref{nd_ni_bif}(a) shows the numerically obtained bifurcation diagram of the system~(\ref{mech}). In Fig.~\ref{nd_ni_bif}(a), $f_0$ is the bifurcation parameter keeping the remaining parameters at fixed values (as written at the beginning of this section). As $f_0$ is increased, a border collision bifurcation occurs where a period-$1$ orbit transforms into other periodic orbits with different periodicities through the chaotic oscillation around the bifurcation. This phenomenon occurs when $m$ is a non-integer one (here, in our case, $m =2.5106$). Fig.~\ref{nd_int_bif}(b) depicts the corresponding maximal Lyapunov Exponent in the parameter space $f_0$. When chaos occurs in the bifurcation diagram, the maximal Lyapunov Exponent shows positive values, and for the periodic orbits, it shows negative. The same phenomenon has been reported earlier in the equivalent one degree of freedom mechanical impacting system \cite{suda2016does}.

\subsection{$m = 2.1$, i.e., close to integer value}
\label{m_int}
\begin{figure}[tbh]
\centering
\begin{subfigure}[b]{\linewidth}
\includegraphics[width = \linewidth]{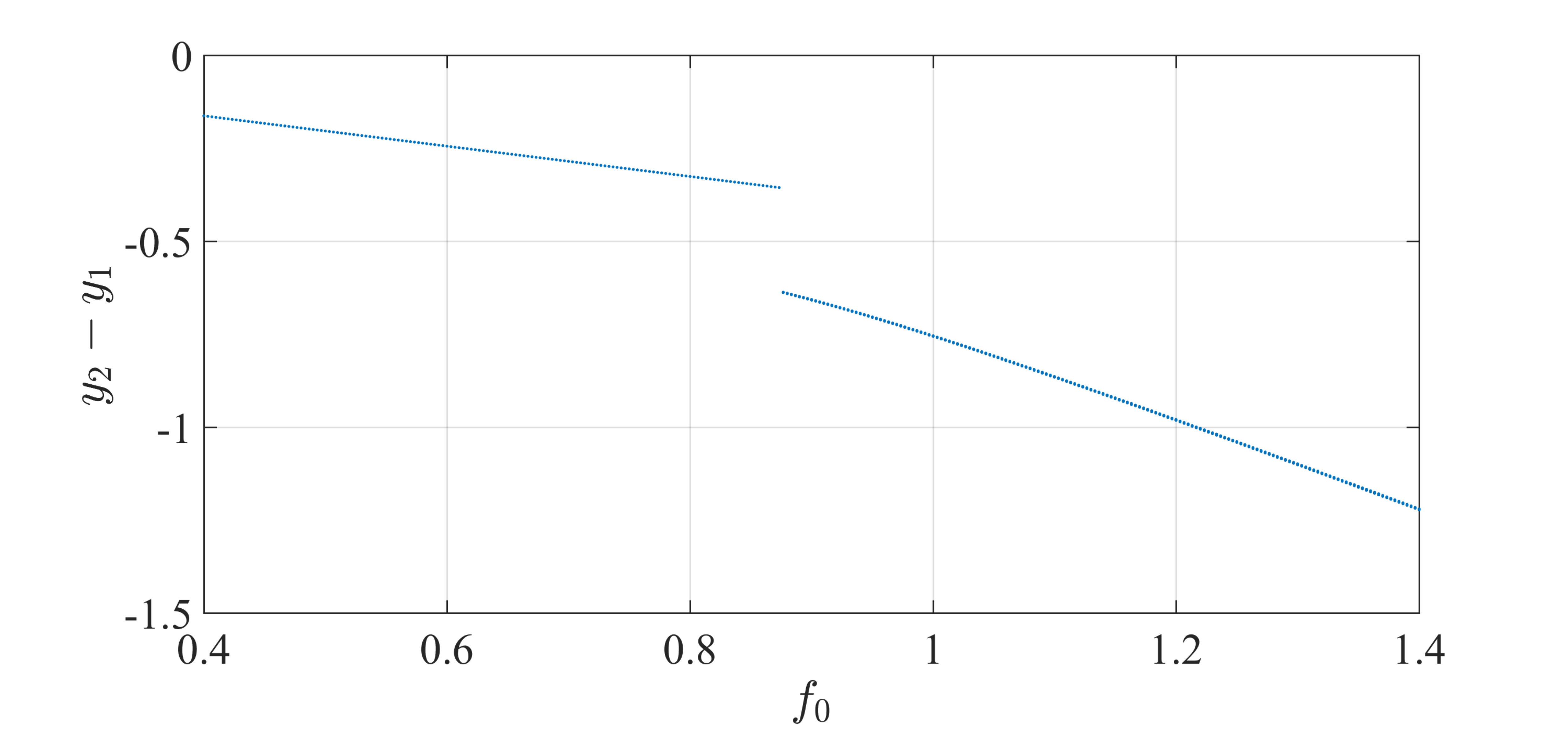}
\caption{}
\end{subfigure}
\begin{subfigure}[b]{\linewidth}
\includegraphics[width = \linewidth]{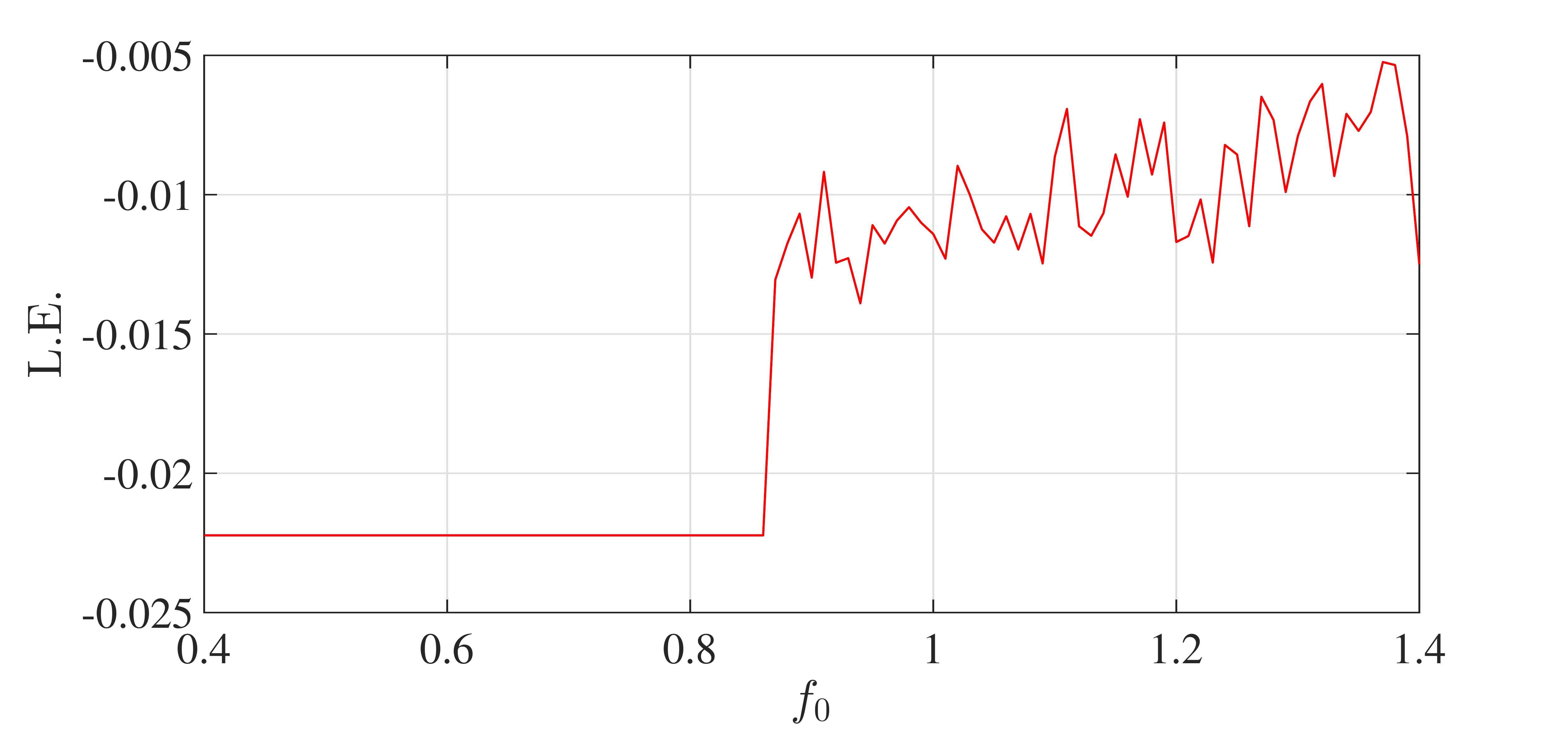}
\caption{}
\end{subfigure}
\caption{(a) Bifurcation diagram and (b) the corresponding maximal Lyapunov Expoent in the non-dimensional parameter space of the system~(\ref{mech}) at close to integer value of $m$. The parameter $f_0$ of both the figures is varied from $0.4$ to $1.4$. (Color Online)}
\label{nd_int_bif}
\end{figure}
Fig.~\ref{nd_int_bif}(a) shows the numerically obtained bifurcation diagram when $m = 2.1$, i.e., close to the integer value. The rest of the parameters have been kept fixed as in the earlier case. When we vary the bifurcation parameter $f_0$ in the forward direction, i.e., from low to high values, a period-$1$ orbit emerges after bifurcation from another period-$1$ orbit. There are no other orbits, for example, a chaotic orbit close to the bifurcation point, which was observed when $m$ is a non-integer one. In the case of Fig.~\ref{nd_int_bif}(a), the bifurcation point is a position where a period-$1$ attractor changes to another period-$1$ attractor having a different amplitude and frequency. This bifurcation diagram confirms that the large amplitude chaos of a two-degree of freedom mechanical impacting system can also be avoided when $m$ is an integer or close to the integer value. The same phenomenon has been reported earlier in the case of a schematic representation of a one-degree of freedom mechanical impacting system \cite{suda2016does}. The corresponding maximal Lyapunov Exponent is shown in Fig.~\ref{nd_int_bif}(b). The negative values of the maximal Lyapunov Exponent throughout the parameter range confirm that the chaotic attractor is absent in the bifurcation diagram.

\begin{figure}[tbh]
  \centering
  \begin{subfigure}[b]{\linewidth}
  \includegraphics[width=\linewidth]{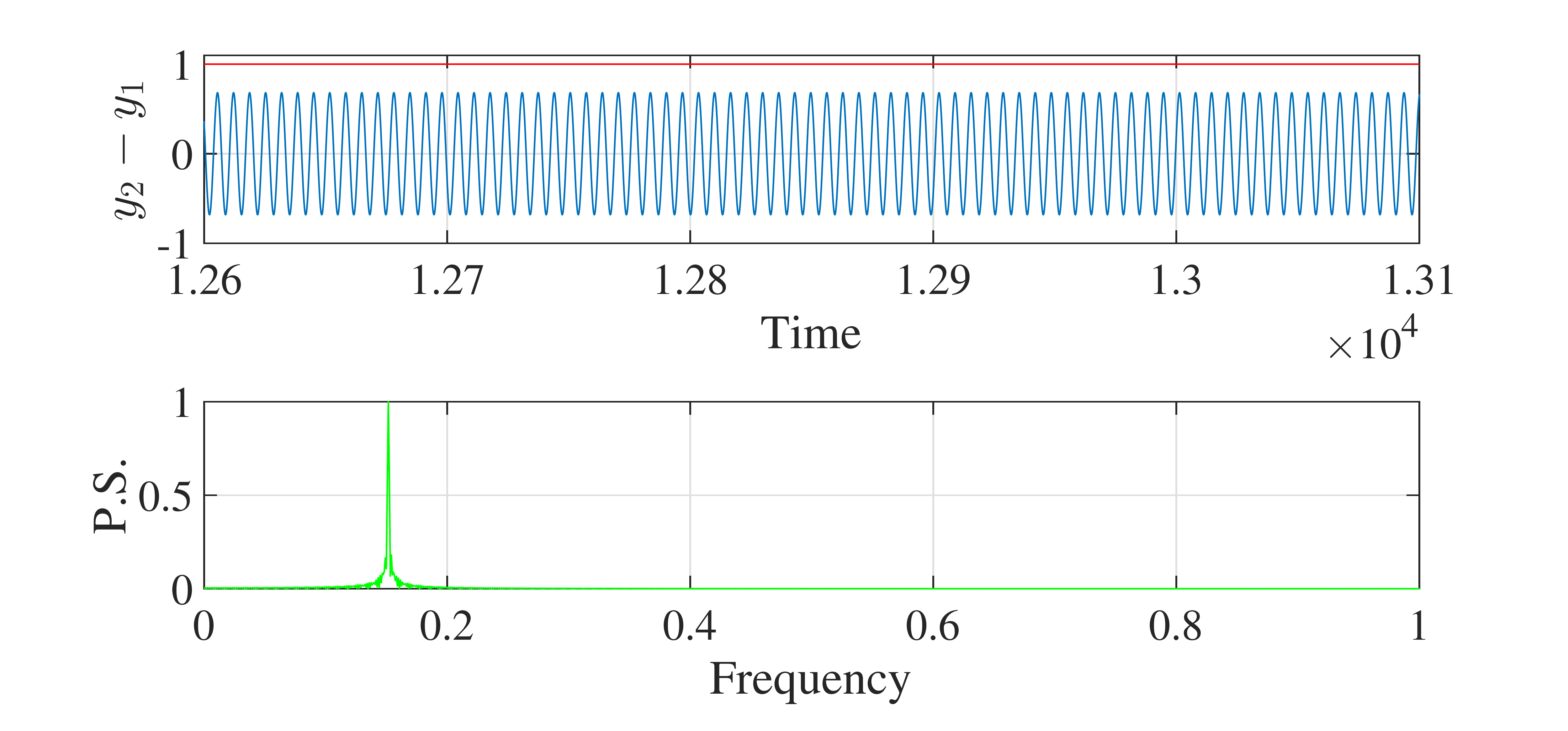}
  \caption{}
  \end{subfigure}  
  \begin{subfigure}[b]{\linewidth}
  \includegraphics[width=\linewidth]{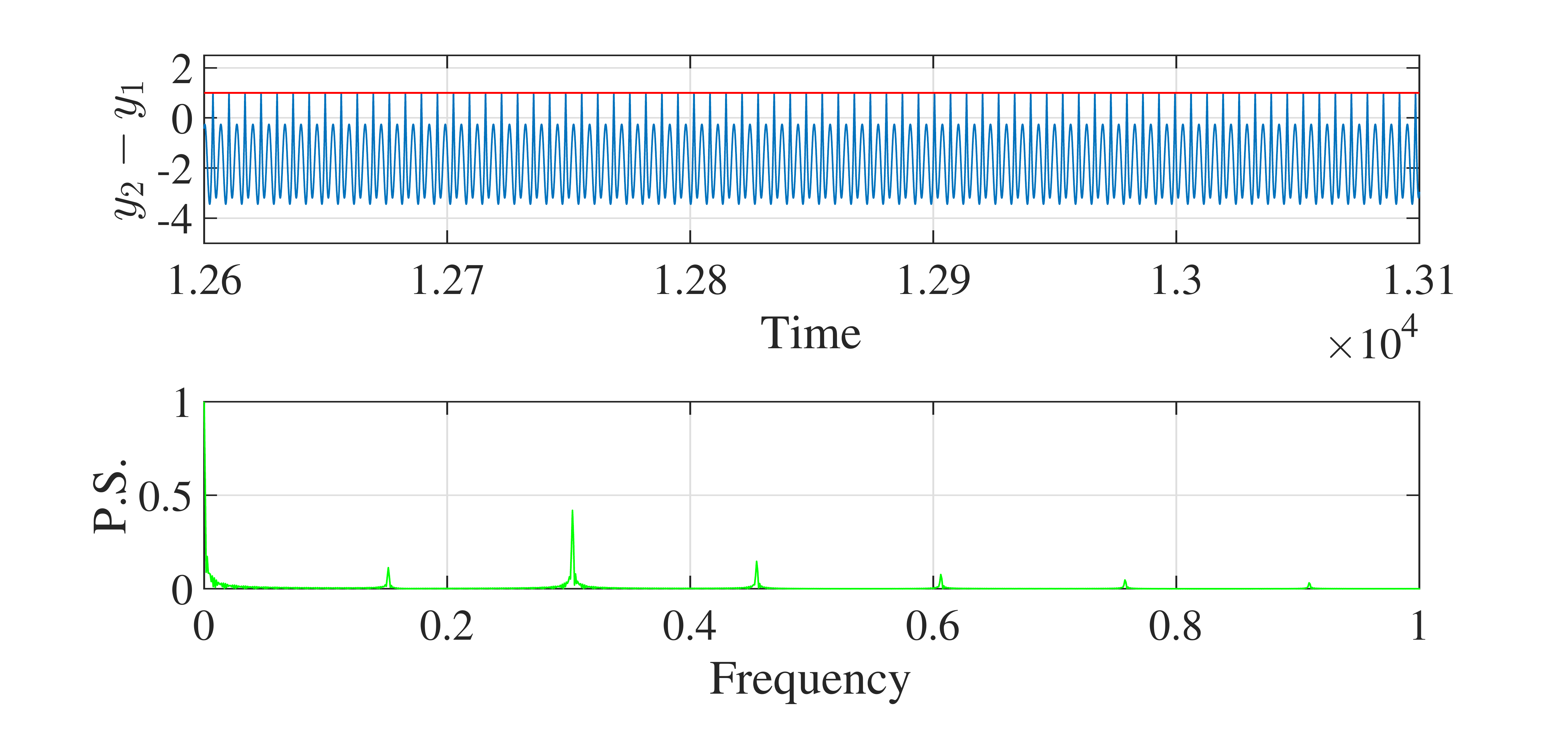}
  \caption{}
  \end{subfigure}   
  \caption{Time-series waveforms and the corresponding frequency spectra for different values of $f_0$ at $m=2.1$. (a) Before impact for $f_0 = 0.6$, Period-$1$ orbit, (b) After the bifurcation for $f_0 = 1.2$, also a period-$2$ orbit. The upper trace of each figure, the $x$-axis is the non-dimensional velocity $y^{\prime}_2$ and the $y$-axes are the non-dimensional displacement $(y_2 - y_1)$ with blue color compared with a constant value $1$ with the red color. In case of the lower trace, the $x$-axis is the nondimensional frequency and the $y$-axis is the normalized power spectra. (Color online.)}
\label{ts_int}
\end{figure}
\begin{figure}[tbh]
  \centering
  \begin{subfigure}[b]{\linewidth}
  \includegraphics[width=\linewidth]{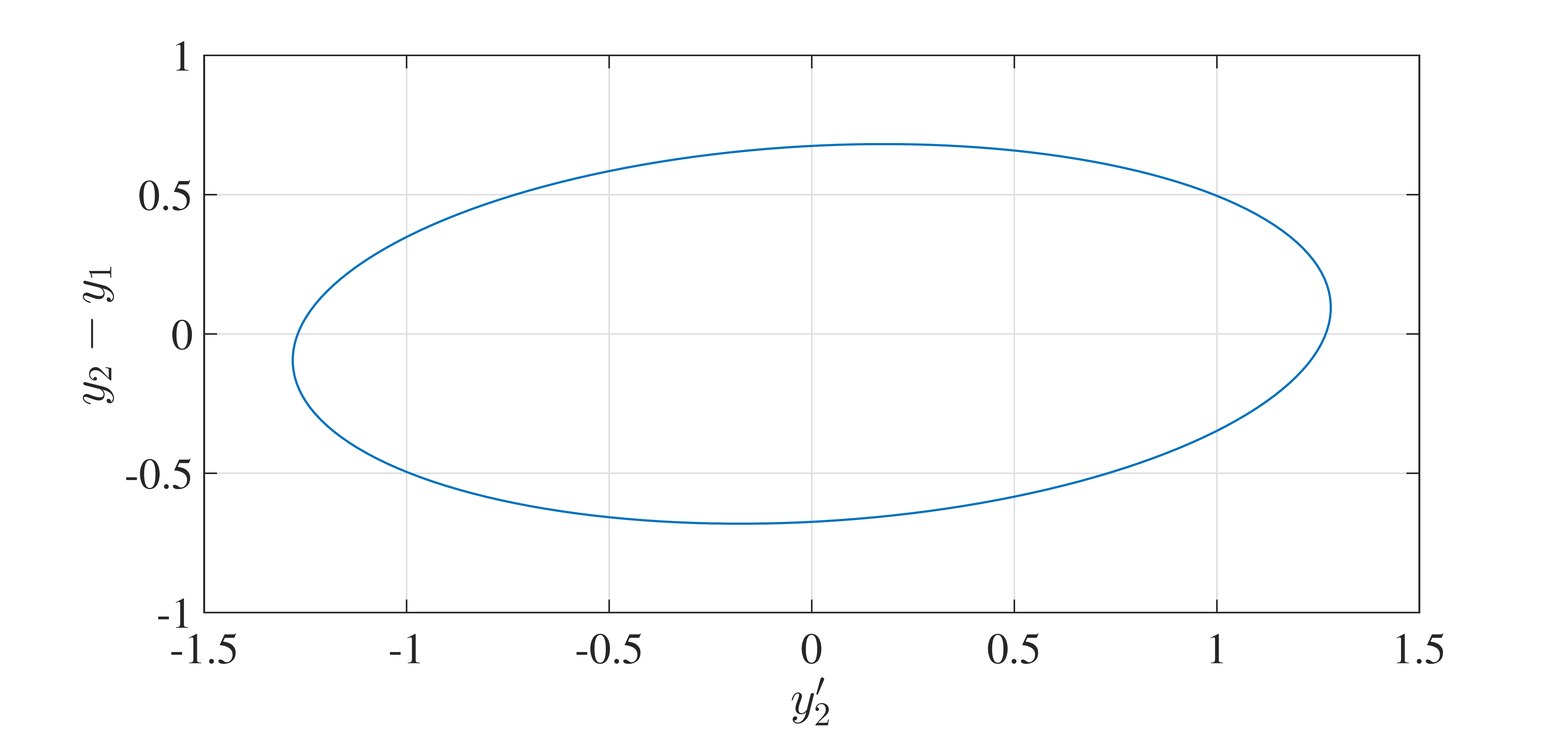}
  \caption{}
  \end{subfigure}  
  \begin{subfigure}[b]{\linewidth}
  \includegraphics[width=\linewidth]{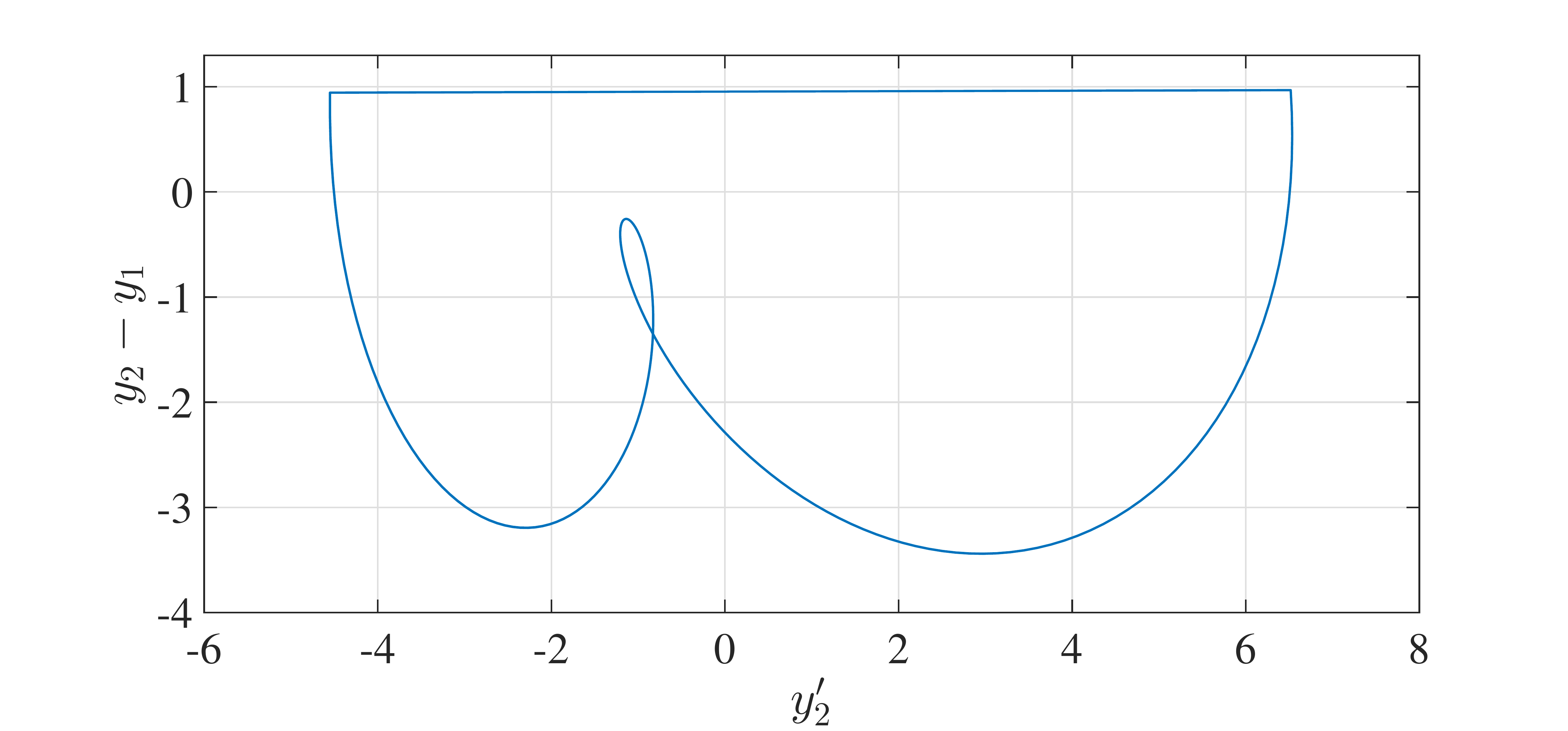}
  \caption{}
  \end{subfigure}   
  \caption{Phase-Space trajectories for different values of $f_0$ at $m=2.1$. (a) Before impact for $f_0 = 0.6$, Period-$1$ orbit, (b) After the bifurcation for $f_0 = 1.2$, also a period-$1$ orbit having different amplitude. The $x$-axis is the non-dimensional velocity $y^{\prime}_2$ and the $y$-axis is the non-dimensional displacement $(y_2 - y_1)$. (Color online.)}
  \label{ps_int}
\end{figure}
Fig.~\ref{ts_int} shows different time-series waveforms of the displacement $(y_2 - y_1)$ compared to a constant $1$ for different values of $f_0$ when $m=2.1$. Fig.~\ref{ts_int}(a) shows the period-$1$ waveform before bifurcation. The waveform of ($y_2-y_1$) does not touch the unit constant value. Fig.~\ref{ps_int}(a) is the corresponding phase-space diagram of the period-$1$ attractor before the bifurcation. The bifurcation happens when $(y_2-y_1)$ touches $1$. Fig.~\ref{ts_int}(b) is the time-series waveform of a period-$2$ attractor after the bifurcation. Fig.~\ref{ps_int}(b) is the corresponding phase-space diagram of period-$2$ orbit after the bifurcation. Here, one loop touches the reference level, i.e., the impacting surface, and the other does not. Please note that while the time series and the phase-space diagrams show the period-$2$ orbits, in the bifurcation diagram, there is one point on the Poincar\'e plane corresponding to that orbit. This mismatch occurs because the considered rig is a non-autonomous dynamical system. So, we have observed the points on the phase-space in the interval of $\tau$, which is the non-dimensional time of the externally applied periodic forcing.

\section{Equivalent electronic circuit of the considered mechanical system}
\label{ckt}
\subsection{Schematic representation}
\label{sr}
\begin{figure}[tbh]
\centering
\includegraphics[width=3in]{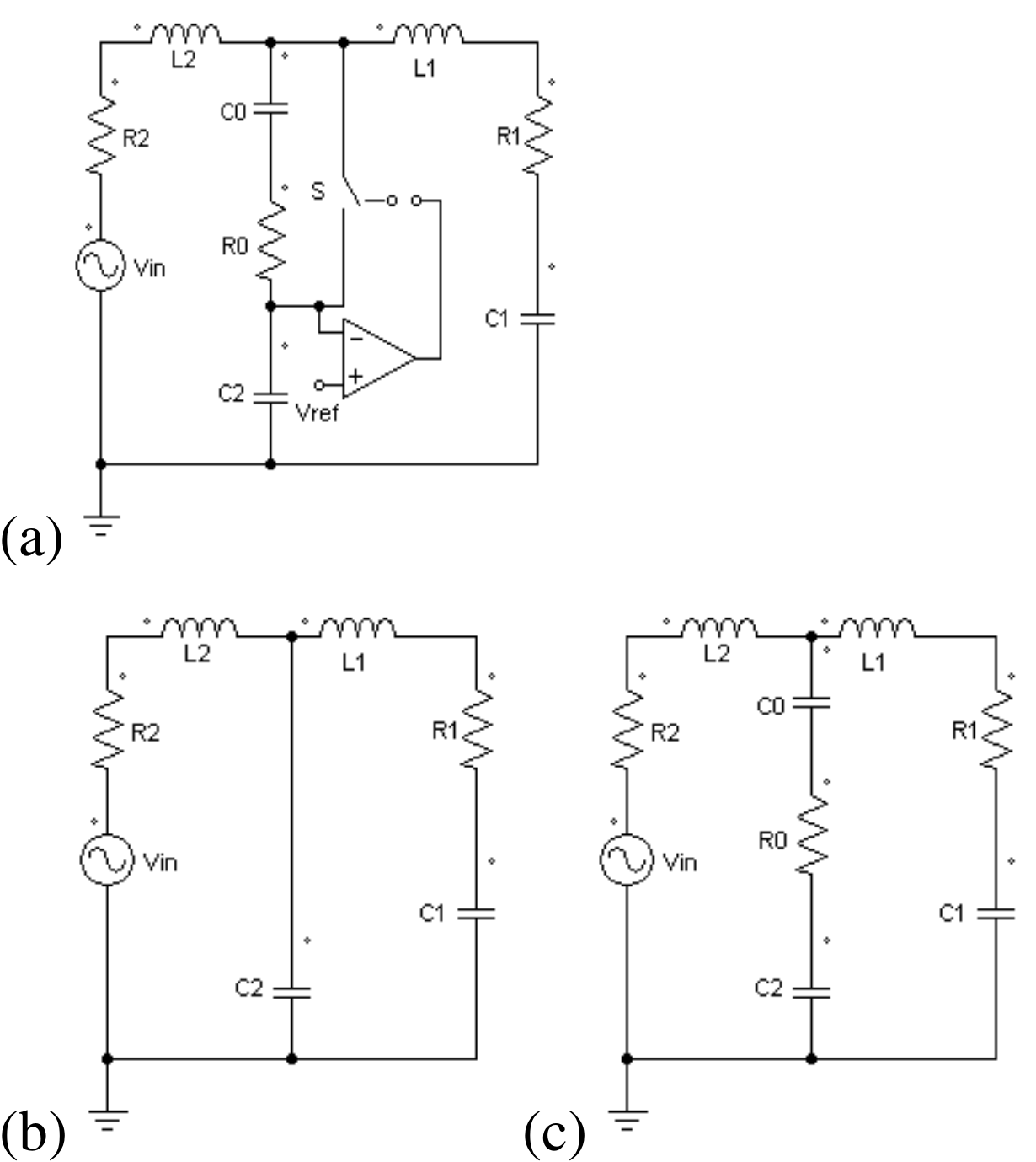}
\caption{(a) The switching circuit under consideration. (b) Switch ON instant: The subsystem for $V_{\rm C2} < V_{\rm ref}$, (c) Switch OFF Instant: The subsystem for $V_{\rm C2} \geq V_{\rm ref}$.}
\label{equiv_ckt}
\end{figure}
The electronic analog circuit of the mechanical system~(\ref{mech}) introduced in this paper is shown in Fig. \ref{equiv_ckt}(a). We have assumed that the equivalent mechanical impacting system has low damping and high stiffness. So, in this case, we have chosen the separation between the impacting wall and the mass $m_1$ is zero. It makes the circuit simpler without changing the dynamics. The electronic circuit system consists of a couple of LCR circuits with an input voltage $V_{\rm in}$ which is a sine wave of amplitude $V_{\rm amp}$ with frequency $f$. The system has an analog switch $S$, controlled by an op-amp-based comparator. When the voltage across the capacitor $C_2$, $V_{\rm C2}$ is less than a reference voltage $V_{\rm ref}$, the switch is `ON', and the circuit configuration is as shown in Fig.~\ref{equiv_ckt}(b). When $V_{\rm C2} \geq V_{\rm ref}$, the switch turns `OFF', connecting the series capacitor $C_0$ and the resistor $R_0$ across the switch with the circuit, which is shown in Fig.~\ref{equiv_ckt}(c).

Let $i_2$ denotes the current passing through the inductor $L_2$ and resistance $R_2$. $i_2 = \frac{dq_2}{dt}$, where, $q_2$ denotes the charge stored in capacitor $C_{\rm 2}$ due to the flow of current $i_2$. Similarly, let $i_1$ is the current passing through the inductor $L_1$, and $i_1 = \frac{dq_1}{dt}$, $q_1$ denotes the charge stored in capacitors due to the current flow $i_1$. These are the state variables of the system. So, the system has the four dimensions, two inductor currents $i_1$, $i_2$, and two electric charges $q_1$, $q_2$.

The parameters are $R_1$, $R_2$, $R_0$, $L_1$, $L_2$, $C_{\rm 1}$, $C_{\rm 2}$, $C_0$, $V_{\rm ref}$, the driving frequency $f = \frac{\Omega_e}{2\pi}$ and $V_{\rm amp}$, where, $\Omega_e$ is the angular frequency of the input sinusoidal signal. Here, $V_{\rm amp}$ has been used as the bifurcation parameters keeping the remaining parameters to constant values.

\subsection{Mathematical model}
\label{math_model}
If we take all the circuit components to be ideal, the system can be described by a set of second order ODEs that form a $4$D piecewise smooth model, given by:

For $(q_2 - q_1) < q_{\rm ref}$,
\begin{equation}
\begin{split}
L_1\ddot{q}_1 &= - R_1\dot{q}_1 - \frac{q_1}{C_1} - \frac{q_1-q_2}{C_2} \\
L_2\ddot{q}_2 & = - R_2\dot{q}_2 -  \frac{q_2 - q_1}{C_2} + V_{\rm amp} \sin(\Omega_e t)
\end{split}
\label{eq1_ckt}
\end{equation}

and, for $(q_2 - q_1) \geq q_{\rm ref}$
\begin{equation}
\begin{split}
L_1\ddot{q}_1 = &- R_1\dot{q}_1 - \frac{q_1}{C_1} -  \frac{q_1 -q_2}{C_2} - R_0 (\dot{q}_1 - \dot{q}_2) \\& - \frac{q_1 - q_2}{C_0} \\
L_2\ddot{q}_2  = &- R_2\dot{q}_2 + \frac{q_1 - q_2}{C_2} + R_0 (\dot{q}_1 - \dot{q}_2) \\& + \frac{q_1 - q_2}{C_0} + V_{\rm amp} \sin(\Omega_e t)
\end{split}
\label{eq2_ckt}
\end{equation}
Here, $q_{\rm ref}$ = $V_{\rm ref}\cdot C_{\rm 2}$. Note that in this system, when the switch is {\sc on} (Fig.\ref{equiv_ckt}b), the system obeys (\ref{eq1_ckt}). Similarly, when the switch is {\sc off} (Fig.\ref{equiv_ckt}c), the system obeys (\ref{eq2_ckt}). The system can be represented in the discrete-time realization by a four-dimensional piecewise smooth map with two compartments divided by a border. The condition represents the borderline between the two compartments that the value of $(q_2 -q_1)$ reaches $q_{\rm ref}$ exactly at the Poincar\'e observation instants (maximum positive amplitude of the input sine wave).

\subsection{Numerical results from the circuit equations}
\label{num_ckt}
We shall show the numerical results obtained from the equations~(\ref{eq1_ckt}) and (\ref{eq2_ckt}) of the equivalent switching electronic circuit as shown in Fig.~\ref{equiv_ckt}(a) to confirm the numerical predictions obtained from the non-dimensional equations of the mechanical system~(\ref{mech}).

\begin{figure}[tbh]
  \centering
  \begin{subfigure}[b]{\linewidth}
  \includegraphics[width=\linewidth]{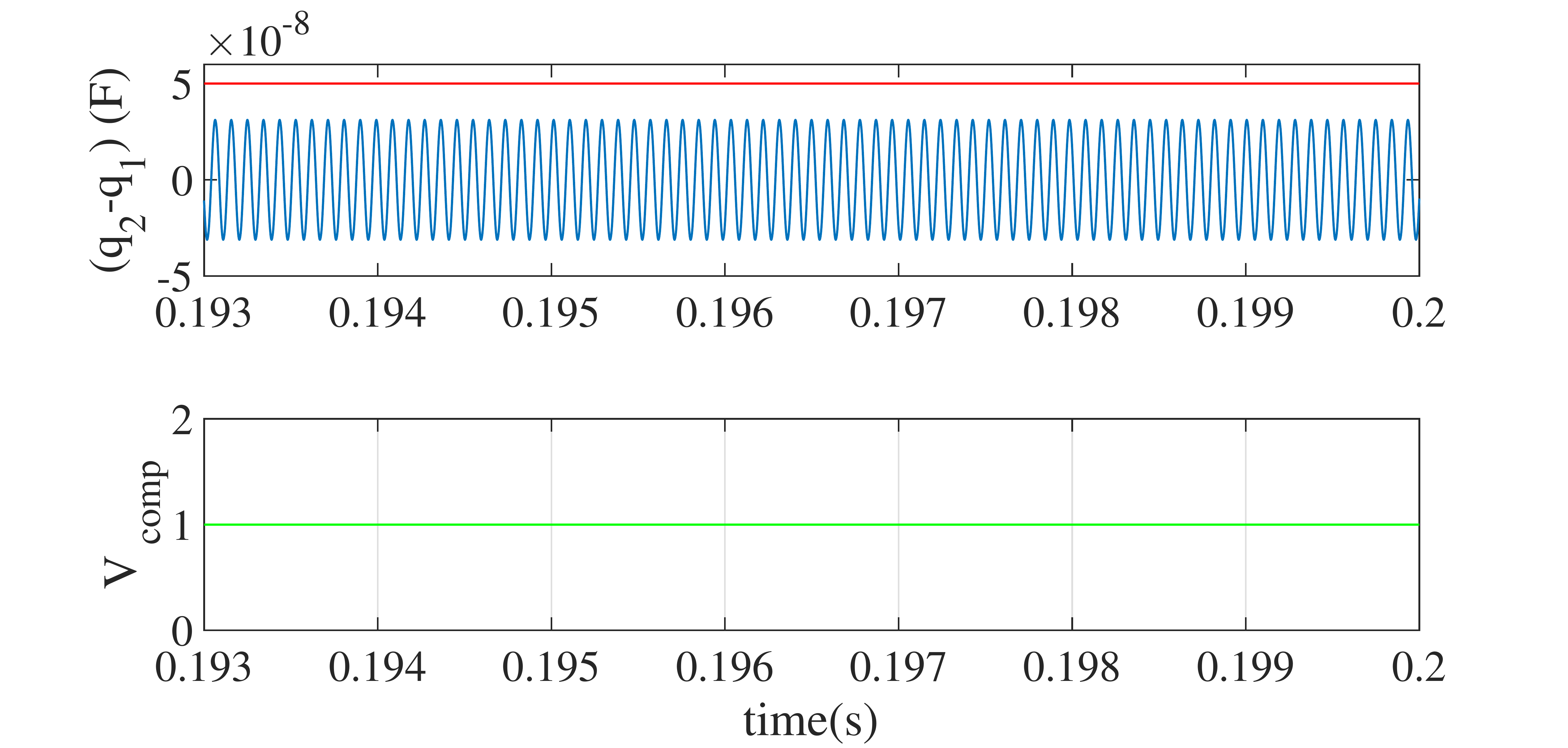}
  \caption{}
  \end{subfigure}  
  \begin{subfigure}[b]{\linewidth}
  \includegraphics[width=\linewidth]{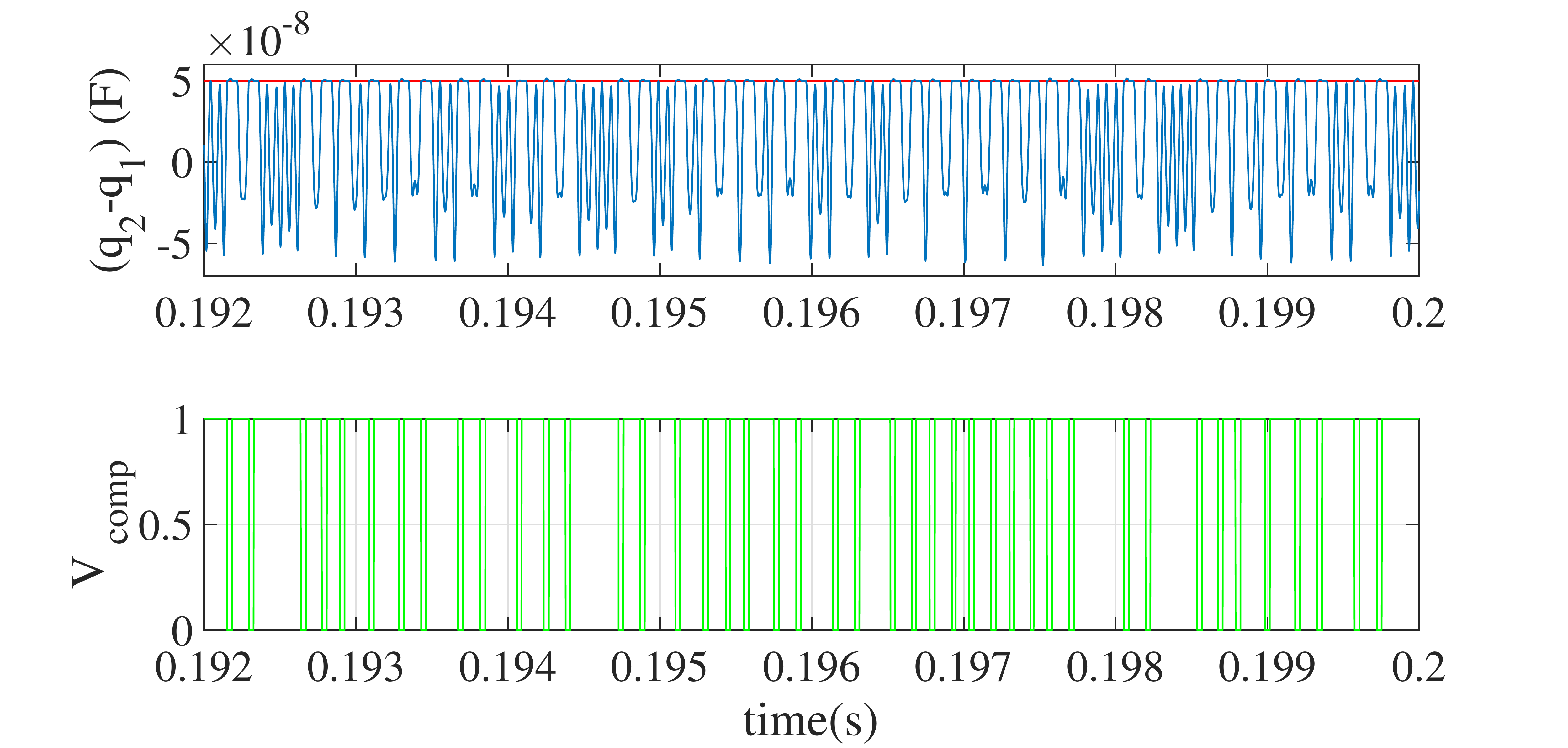}
  \caption{}
  \end{subfigure}
  \begin{subfigure}[b]{\linewidth}
  \includegraphics[width=\linewidth]{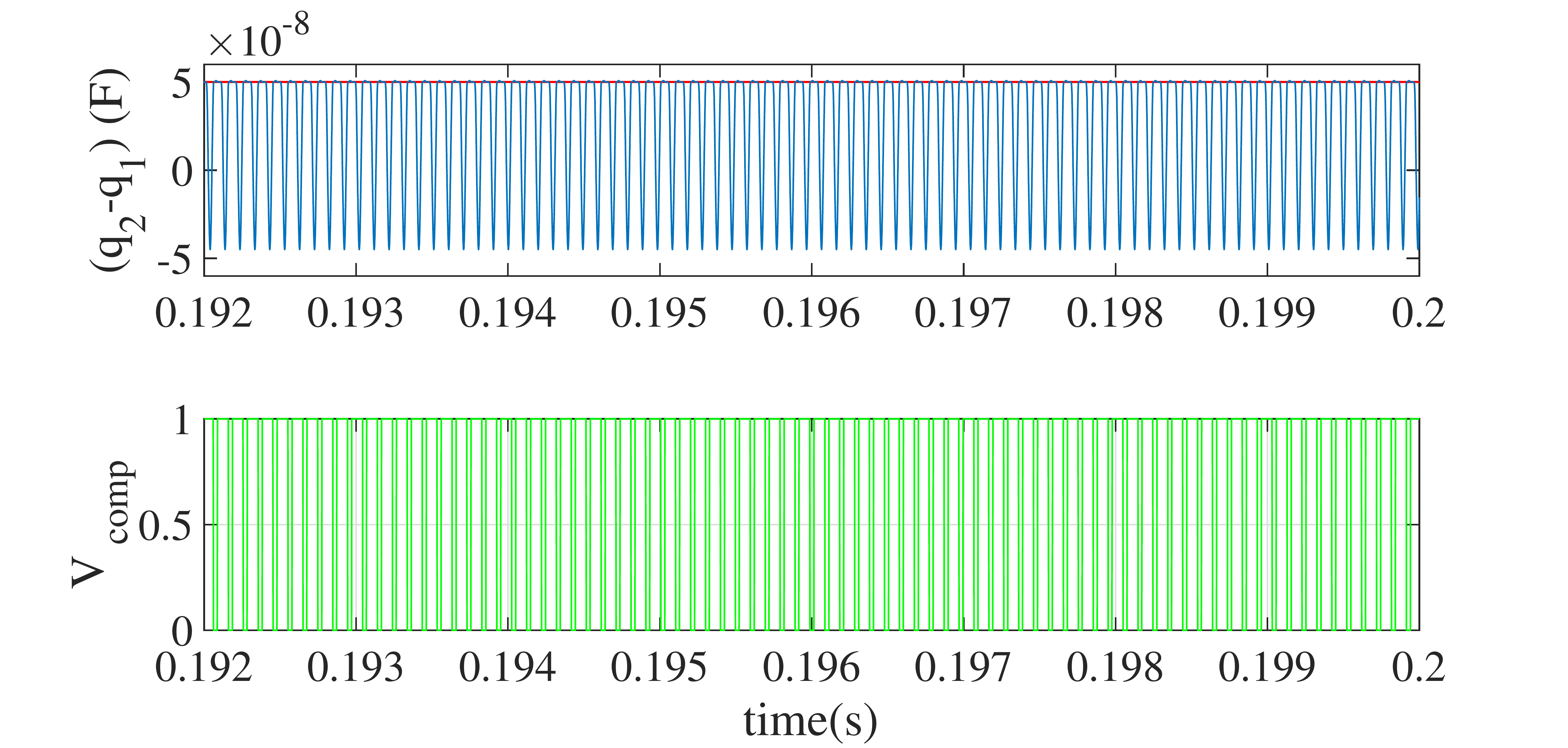}
  \caption{}
  \end{subfigure}
  \caption{Time-Series waveforms of the circuit from simulation. (a) Period-$1$ orbit for $V_{\rm amp} = 0.3$~V, (b) Chaotic orbit for $V_{\rm amp} = 0.57$~V, (c) Period-$1$ orbit at $V_{\rm amp} = 0.71$~V. In each subfigure, the upper trace is the charge stored in capacitor $C_2$ compared with $q_{\rm ref} = V_{\rm ref} \cdot C_2$. and the lower trace is the normalized comparator output. The initial condition is chosen at $(-0.5V,-0.1V,-0.01V,1V,0.001V)$. (Color online.)}
\label{ts_ckt}
\end{figure}
Fig.~\ref{ts_ckt} shows a few numerically obtained time-series waveforms for the considered equivalent circuit. The parameter values are: $L_1 = 100.4$~mH, $L_2 = 10.21$~mH, $R_1 = 230$~$\Omega$, $R_2 = 29.7$~$\Omega$, $C_1 = 1.601$~nF, $C_2 = 16.680$~nF, $R_0 = 10.5$~$\Omega$, $C_0 = 64.0$~pF, $V_{\rm ref} = 3.0$~V, and $f = 10$~kHz. The amplitude of the applied sine voltage, $V_{\rm amp}$, is to be varied to obtain different dynamical behaviors. When the charge stored in the capacitor, $(q_2-q_1)$ does not exceed the reference charge $q_{\rm ref}$, the comparator output is high, i.e., $1$. This is the condition before border collision, and the circuit shows a period-$1$ waveform, shown in Fig.~\ref{ts_ckt}(a). When $V_{\rm amp}$ is increased to such a value that $(q_2-q_1)$ just touches $q_{\rm ref}$, at the point of bifurcation, the dynamics of the state-variable becomes chaotic in nature (as shown in Fig.~\ref{ts_ckt}(b)). The comparator output in that condition becomes erratic. For the higher value of $V_{\rm amp}$, $(q_2-q_1)$ becomes again periodic. The corresponding time-series waveform and the comparator output are shown in Fig.~\ref{ts_ckt}(c). So, we can find that the system shows the chaotic attractor for a small range in the parameter space, i.e., during the grazing condition. Before and after the grazing conditions, the orbits are periodic.

\begin{figure}[tbh]
  \centering
  \begin{subfigure}[b]{\linewidth}
  \includegraphics[width=\linewidth]{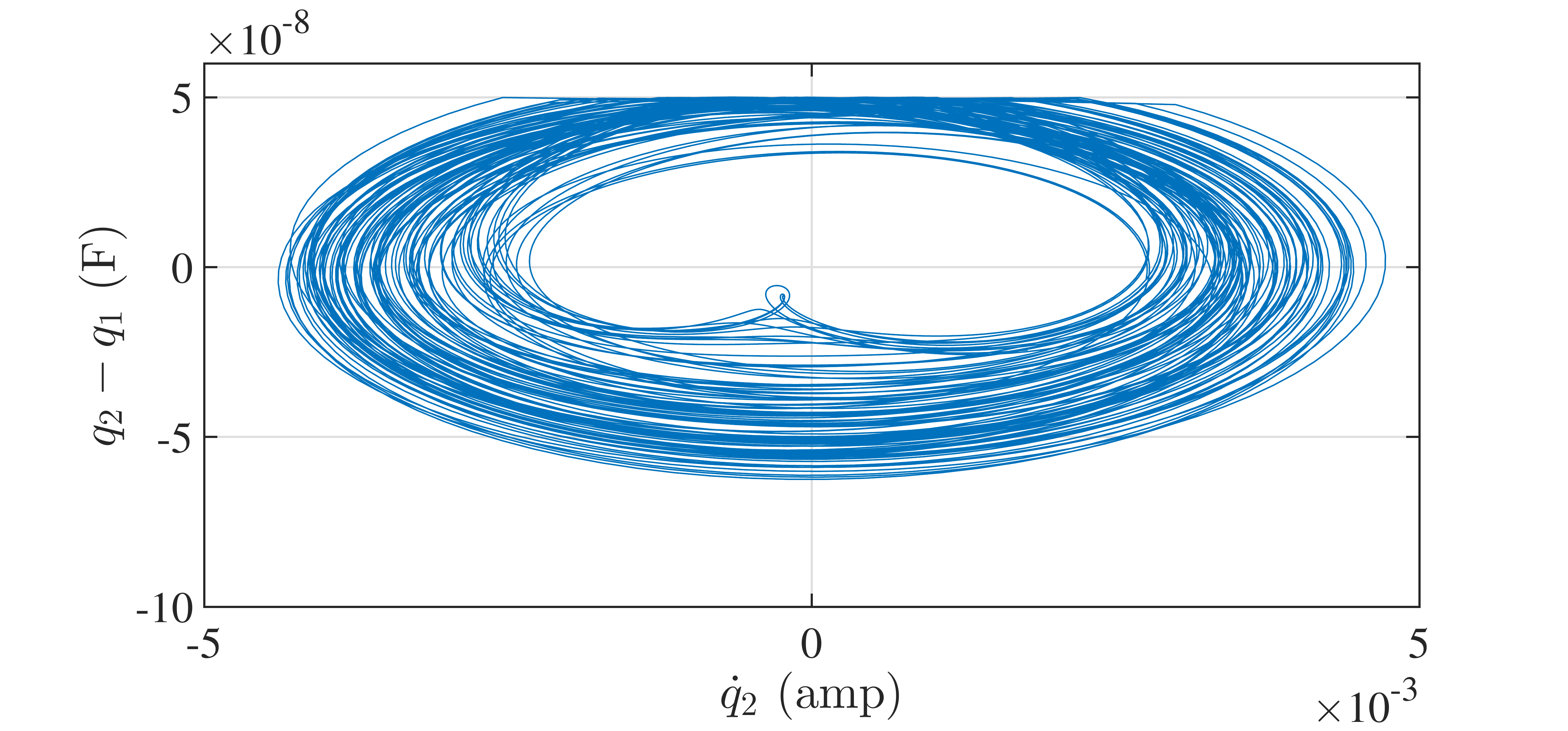}
  \caption{}
  \end{subfigure}  
  \begin{subfigure}[b]{\linewidth}
  \includegraphics[width=\linewidth]{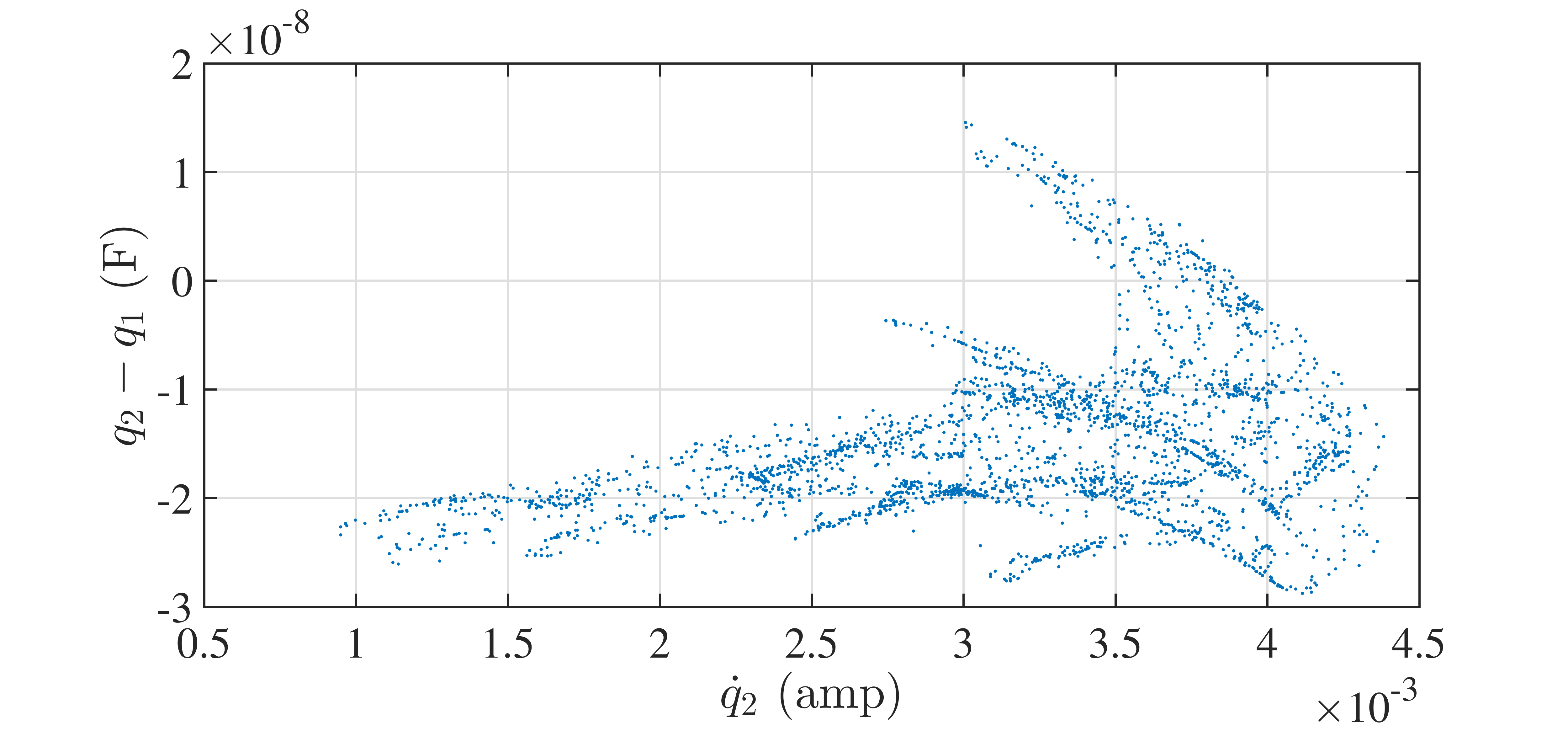}
  \caption{}
  \end{subfigure}
  \caption{(a) Phase space diagram and (b) Poincar\'e section of the chaotic attractor at grazing for $m$ non-integer condition. The $x$-axis is the value of the current flowing through the inductor $L_2$ and the $y$-axis is the charge of the capacitor $C_2$, $(q_2-q_1)$, in F. (Color online.)}
\label{ps_chaos}
\end{figure}
Fig.~\ref{ps_chaos}(a) shows the phase-space diagram of the circuit for $m$ non-integer condition. The amplitude of the externally applied sine wave is $V_{\rm amp} = 0.57$~V. The remaining fixed parameter is shown earlier in this section. The attractor is chaotic during the bifurcation. Fig.~\ref{ps_chaos}(b) depicts the Poincar\'e section of that chaotic attractor at grazing. Here also, the discrete-time representation of the chaotic attractor is finger-shaped. This finger-shaped attractor was discussed earlier in \cite{banerjee2009invisible}.

\begin{figure}[tbh]
  \centering
  \begin{subfigure}[b]{\linewidth}
  \includegraphics[width=\linewidth]{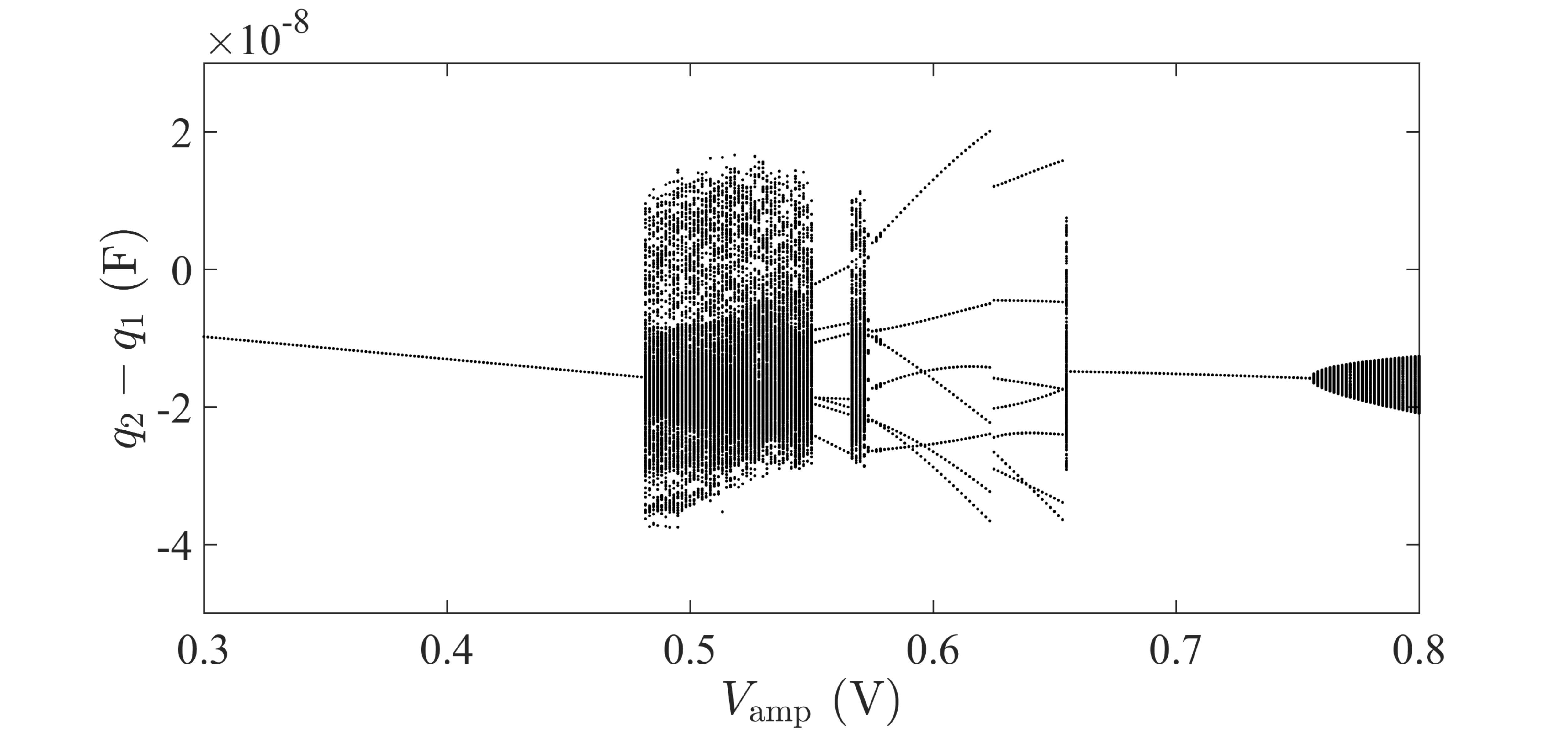}
  \caption{}
  \end{subfigure}  
  \begin{subfigure}[b]{\linewidth}
  \includegraphics[width=\linewidth]{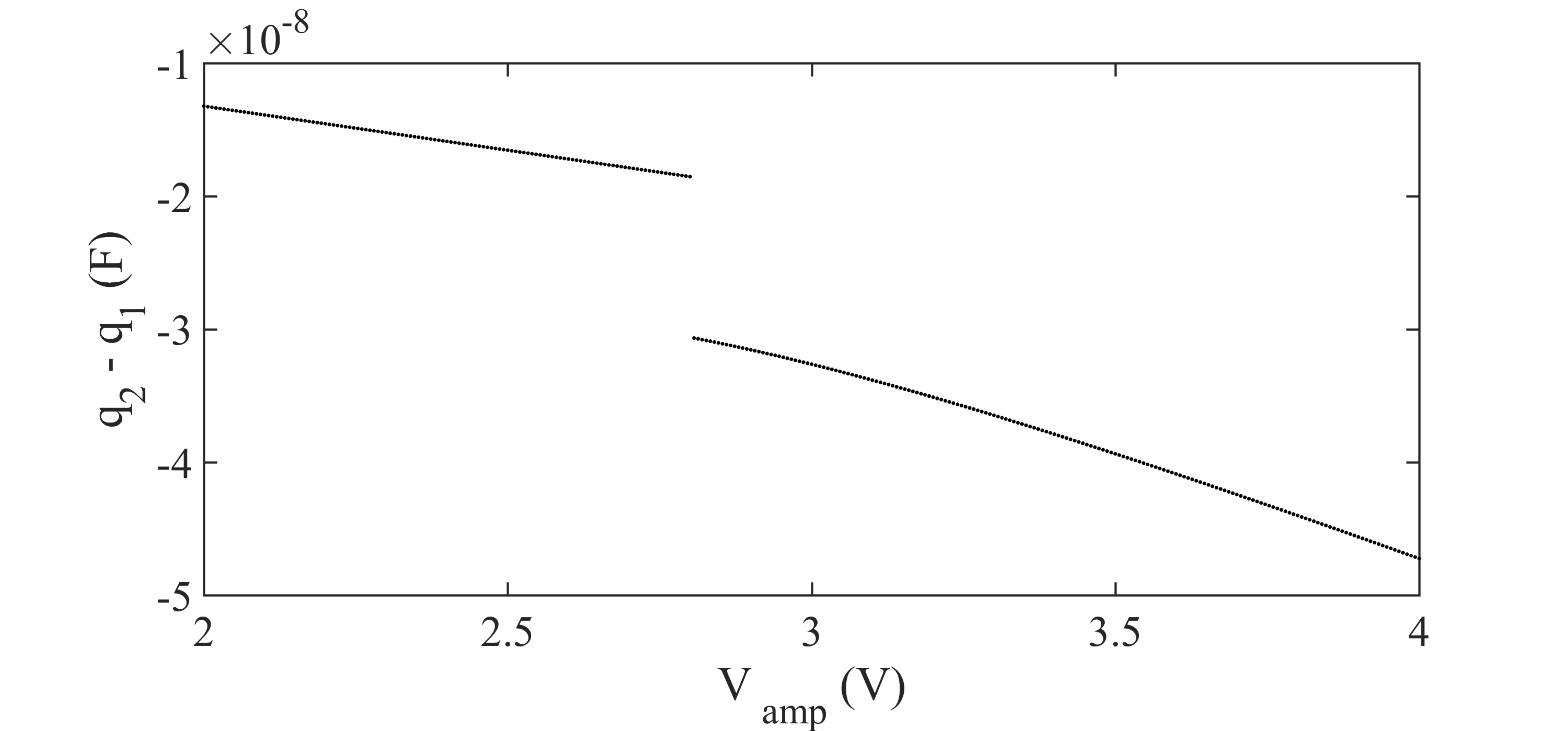}
  \caption{}
  \end{subfigure}
  \caption{Bifurcation diagrams of the equivalent circuit when input voltage amplitude $V_{\rm amp}$ is used as varying parameter for (a) $m$ is non-integer and (b) $m$ is close to an integer value. For (a) $f = 10$~kHz and for (b) $f = 11.96$~kHz. The $x$-axis is the amplitude of the externally applied sine wave, $V_{\rm amp}$ in V and the $y$-axis is the charge of the capacitor $C_2$ in F.}
\label{bif_ckt}
\end{figure}
The bifurcation diagram of the considered circuit is shown in Fig.~\ref{bif_ckt}(a) when $m$ is a non-integer value. $V_{\rm amp}$ is taken as the bifurcation parameter, keeping the remaining parameters fixed (as stated earlier in this section). When $m$ is a non-integer, the diagram exhibits a border collision bifurcation from a periodic orbit to another periodic orbit with different periodicities through a large amplitude chaotic oscillation around the bifurcation point. The same thing has happened in Fig.~\ref{nd_ni_bif}. Fig.~\ref{bif_ckt}(b) depicts the bifurcation diagram when $m$ is close to integer value. There, we cannot find any chaotic attractor at the bifurcation. The same phenomena were observed for the equivalent circuit of a one-degree of freedom mechanical impacting system in \cite{seth2020electronic}.

\section{Experimental results}
\label{expt_res}
\begin{figure}[tbh]
\centering
\includegraphics[width=3in]{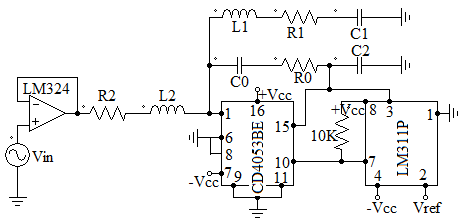}
\caption{Circuit diagram of the experimental system. LM324 is an Op-Amp, LM311P is a comparator and CD4053BE is a CMOS single 8-channel analog multiplexer/demultiplexer with logic-level conversion. The parameter values are: $L_1 = 100.4$~mH (internal impedence $71.7$~$\Omega$), $L_2 = 10.21$~mH (internal impedence $13.6$~$\Omega$), $R_1 = 158.3$~$\Omega$, $R_2 = 16.1$~$\Omega$, $C_1 = 1.601$~nF, $C_2 = 16.680$~nF, $R_0 = 10.5$~$\Omega$, $C_0 = 64.0$~pF, $f = 10$~kHz, and $V_{\rm ref} = 3.0$~V. The supply voltages have been taken as $\pm V_{\rm CC} = \pm 12$~V. (Color Online.)}
\label{ckt_impl}
\end{figure}
In order to validate the numerically predicted results, we have developed an experimental setup with the same parameter values. Fig.~\ref{ckt_impl} shows the circuit diagram used for implementation. Here, we have used a quad LM$324$ op-amp as a unity gain buffer to minimize the signal generator's loading effect. The LM$311$P is used as a comparator, which compares $V_{\rm C2}$, i.e., the voltage across the capacitor $C_2$, with the reference constant voltage $V_{\rm ref}$. When $V_{\rm C2} < V_{\rm ref}$, the comparator stays in the {\sc on} state, else the output of the comparator is in the {\sc off} state. CD$4053$BE is used as the analog switch. The input voltage's amplitude value, $V_{\rm in}$, is chosen from a signal generator.

\begin{figure}[tbh]
  \centering
  \begin{subfigure}[b]{0.49\linewidth}
  \includegraphics[width=\linewidth]{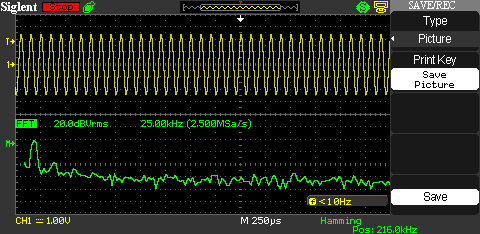}
  \caption{}
  \end{subfigure}  
  \begin{subfigure}[b]{0.49\linewidth}
  \includegraphics[width=\linewidth]{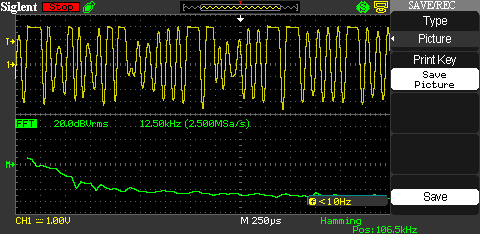}
  \caption{}
  \end{subfigure}
  
  \begin{subfigure}[b]{0.5\linewidth}
  \includegraphics[width=\linewidth]{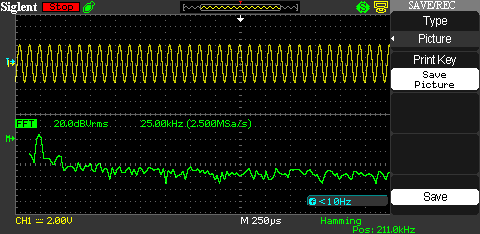}
  \caption{}
  \end{subfigure}
  
  \caption{Time-Series waveforms of the circuit from experiment: $x$-axis is the time and $y$-axis is the voltage in V. (a) Period-$1$ waveform for $V_{\rm amp} = 0.3$~V (Grid division: $1.00$~V), (b) Chaotic waveform for $V_{\rm amp} = 0.57$~V (Grid division: $1.00$~V), (c) Period-$1$ waveform after the bifurcation at $V_{\rm amp} = 0.72$~V (Grid division: $2.00$~V). In each subfigure, the upper trace is the voltage across the capacitor $C_2$. the lower trace is the Spread
spectrum characteristics. FFT sample rate $2.50$~MSa/s, span $25.00$~kHz, center $216.0$~kHz,
scale $20$~dB.  (Color online.)}
\label{expt_ts}
\end{figure}
Fig.~\ref{expt_ts} shows the time-series waveforms obtained experimentally of the equivalent circuit~(\ref{equiv_ckt}). At $V_{\rm amp}$ = $0.3$~V, there is a period-$1$ orbit as shown in Fig.~\ref{expt_ts}(a). The single peak in the frequency spectrum confirms the orbit is periodic. When the parameter $V_{\rm amp}$ is varied, a chaotic waveform generates at $V_{\rm amp} = 0.57$~V. The corresponding frequency spectra confirm the orbit is chaotic, as shown in Fig.~\ref{expt_ts}(b). This is the condition during the border collision bifurcation. After the bifurcation, the voltage across $C_2$, $V_{\rm C2}$, becomes periodic which is shown in Fig.~\ref{expt_ts}(c). The corresponding frequency spectra are shown with the waveform.

\begin{figure}[tbh]
  \centering
  \begin{subfigure}[b]{0.49\linewidth}
  \includegraphics[width=\linewidth]{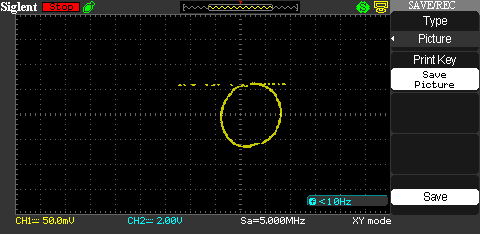}
  \caption{}
  \end{subfigure}  
  \begin{subfigure}[b]{0.49\linewidth}
  \includegraphics[width=\linewidth]{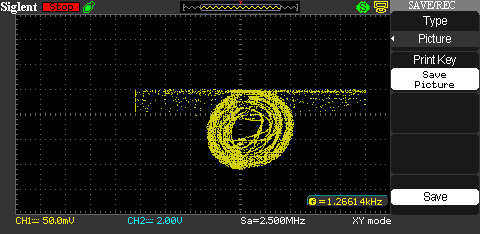}
  \caption{}
  \end{subfigure}
  
  \begin{subfigure}[b]{0.5\linewidth}
  \includegraphics[width=\linewidth]{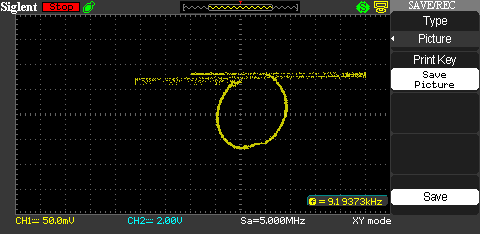}
  \caption{}
  \end{subfigure}
  
  \caption{Phase-space diagrams for different attractors obtained from experiment: $x$-axis is the voltage across resistance $R_2$, i.e., $V_{\rm R2}$ in V and $y$-axis is the voltage across the capacitor $C_2$, $V_{\rm C2}$ in V. (a) Period-$1$ orbit for $V_{\rm amp} = 0.3$~V, (b) Chaotic orbit for $V_{\rm amp} = 0.57$~V, (c) Period-$1$ orbit after bifurcation at $V_{\rm amp} = 0.72$~V. In each subfigure, the grid divisions are: along the $x$-axis $50$~mV and along the $y$-axis $2.00$~V. (Color online.)}
\label{expt_ps}
\end{figure}
The experimentally obtained phase-space trajectories for different values of $V_{\rm amp}$ are shown in Fig.~\ref{expt_ps}. Fig.~\ref{expt_ps}(a) shows the period-$1$ orbit for $V_{\rm amp} = 0.3$~V. The single loop in the state-space confirms that the orbit is periodic. This is the before bifurcation condition. In Fig.~\ref{expt_ps}(b), the erratic nature of the orbit in the phase-space makes the system chaotic at $V_{\rm amp} = 0.57$~V. This is the condition during bifurcation. Fig.~\ref{expt_ps}(c) shows the existence of period-$1$ attractor after the bifurcation.

So, from the experiment also, we can say that when the forced frequency is a non-integer multiple of twice the average value of the two natural frequencies, we get chaos in the bifurcation diagram around the grazing of the equivalent circuit of the mechanical oscillator.

\begin{figure}[tbh]
  \centering
  \begin{subfigure}[b]{0.49\linewidth}
  \includegraphics[width=\linewidth]{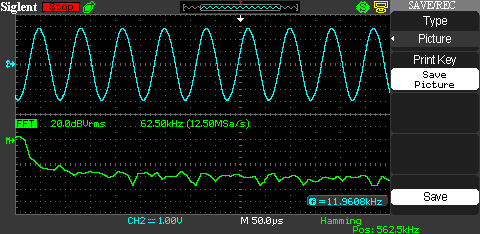}
  \caption{}
  \end{subfigure}  
  \begin{subfigure}[b]{0.49\linewidth}
  \includegraphics[width=\linewidth]{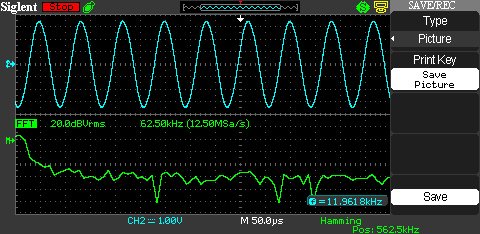}
  \caption{}
  \end{subfigure}
  
  \caption{Time-Series waveforms of the circuit from experiment: $x$-axis is the time and $y$-axis is the voltage in V. (a) Period-$1$ waveform for $V_{\rm amp} = 2.50$~V (Grid division: $1.00$~V), (b) Period-$1$ waveform after the bifurcation at $V_{\rm amp} = 3.10$~V (Grid division: $1.00$~V). In each subfigure, the upper trace is the voltage across the capacitor $C_2$. the lower trace is the Spread
spectrum characteristics. FFT sample rate $2.50$~MSa/s, span $25.00$~kHz, center $216.0$~kHz,
scale $20$~dB.  (Color online.)}
\label{expt_int_ts}
\end{figure}
Fig.~\ref{expt_int_ts} depicts the experimentally obtained time-series waveforms when the externally applied frequency is close to the integer multiple of twice the average of the natural frequencies. Fig.~\ref{expt_int_ts}(a) shows the period-$1$ waveform before bifurcation at $V_{\rm amp} = 2.50$~V. The single peak in the frequency spectra confirms that the waveform is periodic. At that time, the maximum amplitude of the $V_{\rm C2}$ does not cross the DC reference voltage, i.e., $V_{\rm ref} = 3.0$~V. Fig.~\ref{expt_int_ts}(b) signifies the waveform of $V_{\rm C2}$ after the bifurcation at $V_{\rm amp} = 3.10$~V. This waveform is also period-$1$ in nature. At that time, the voltage $V_{\rm C2}$ has crossed the DC voltage $V_{\rm ref} = 3.0$~V. If one can notice carefully, we can find that the time-series waveform after the bifurcation at $V_{\rm amp} = 3.10$~V, obtained experimentally, is not the same as the numerical prediction. The reason behind it is that in the case of the experimental circuit implementation, there are some ambiguities we do not have any control over, like the resistance of the connecting wires, the tolerance level of the capacitors and inductors, ICs time delay of propagation, etc. Thus, the waveform has changed from the numerical predictions. But as a whole, the dynamics remain the same. We have also checked and confirmed the other parameter values in the bifurcation diagram. We have not obtained any chaotic waveform in the whole parameter range in the bifurcation diagram. The chaotic attractor is absent in the bifurcation diagram, confirmed during the experiment. So, the experimental results confirm that the chaotic oscillation can be avoided when the $m$ is either integer or close to the integer one. The experimental results validate the results predicted numerically.

\section{Conclusions}
In this paper, we have proposed a two-degrees of freedom forced damped mechanical impacting oscillator. We have shown that when the externally applied frequency is not an integer multiple of the addition of the two natural frequencies, there is an onset of chaos in the bifurcation diagram. The chaotic attractor at the bifurcation point has a typical finger-shaped in discrete-time representation. A similar situation has occurred for a one-degree of freedom mechanical impacting system. Also, we have shown that if the forcing frequency is close to the integer times the summation of the two natural frequencies, the onset of chaos can be avoided.

We have developed the non-dimensionalized equations of the considered system. We have shown that the system exhibits the same dynamical phenomena reported in the case of one-degree of freedom mechanical impact oscillators. This includes the occurrence of narrowband chaos, a finger-shaped attractor occurring close to the grazing parameter value, the disappearance of the chaotic oscillation at specific parameter ranges, etc.

We have introduced an electronic analogue of the considered mechanical impacting system. It provides a straightforward and convenient experimental platform with which various practical explorations on impacting systems can be carried out. We have confirmed the numerical predictions using the actual circuit results.
 
\begin{acknowledgements}
This work has been supported by the Polish National Science Centre, Poland under the grant OPUS 18 No. 2019/35/B/ST8/00980.

\end{acknowledgements}

\section*{Data availability}
Data sharing is not applicable to this article.

\section*{Declarations}

 \section*{Conflict of interest}
 The authors declare that they have no conflict of interest.

\bibliographystyle{spmpsci}       % APS-like style for physics
\bibliography{mybibfile}   % name your BibTeX data base

\begin{thebibliography}{10}
\providecommand{\url}[1]{{#1}}
\providecommand{\urlprefix}{URL }
\expandafter\ifx\csname urlstyle\endcsname\relax
  \providecommand{\doi}[1]{DOI~\discretionary{}{}{}#1}\else
  \providecommand{\doi}{DOI~\discretionary{}{}{}\begingroup
  \urlstyle{rm}\Url}\fi

\bibitem{awrejcewicz2002nonlinear}
Awrejcewicz, J., Kudra, G., Lamarque, C.H.: Nonlinear dynamics of triple
  pendulum with impacts.
\newblock Journal of Technical Physics \textbf{Vol. 43, no 2}, 97--112 (2002)

\bibitem{awrejcewicz2003bifurcation}
Awrejcewicz, J., Lamarque, C.: Bifurcation And Chaos In Nonsmooth Mechanical
  Systems.
\newblock World Scientific Series On Nonlinear Science Series A. World
  Scientific Publishing Company (2003).
\newblock \urlprefix\url{https://books.google.pl/books?id=WxnJCgAAQBAJ}

\bibitem{banerjee2009invisible}
Banerjee, S., Ing, J., Pavlovskaia, E., Wiercigroch, M., Reddy, R.K.: Invisible
  grazings and dangerous bifurcations in impacting systems: The problem of
  narrow-band chaos.
\newblock Phys. Rev. E \textbf{79}, 037201 (2009).
\newblock \doi{10.1103/PhysRevE.79.037201}.
\newblock \urlprefix\url{https://link.aps.org/doi/10.1103/PhysRevE.79.037201}

\bibitem{blazejczyk1998co}
Blażejczyk-Okolewska, B., Kapitaniak, T.: Co-existing attractors of impact
  oscillator.
\newblock Chaos, Solitons \& Fractals \textbf{9}(8), 1439--1443 (1998).
\newblock \doi{https://doi.org/10.1016/S0960-0779(98)00164-7}.
\newblock
  \urlprefix\url{https://www.sciencedirect.com/science/article/pii/S0960077998001647}

\bibitem{budd1995grazing}
Budd, C.: Grazing in Impact Oscillators, pp. 47--63.
\newblock Springer Netherlands, Dordrecht (1995).
\newblock \doi{10.1007/978-94-015-8439-5_3}.
\newblock \urlprefix\url{https://doi.org/10.1007/978-94-015-8439-5_3}

\bibitem{dankowicz2000origin}
Dankowicz, H., Nordmark, A.B.: On the origin and bifurcations of stick-slip
  oscillations.
\newblock Physica D: Nonlinear Phenomena \textbf{136}(3), 280--302 (2000).
\newblock \doi{https://doi.org/10.1016/S0167-2789(99)00161-X}.
\newblock
  \urlprefix\url{https://www.sciencedirect.com/science/article/pii/S016727899900161X}

\bibitem{di2001normal}
{di Bernardo}, M., Budd, C., Champneys, A.: Normal form maps for grazing
  bifurcations in n-dimensional piecewise-smooth dynamical systems.
\newblock Physica D: Nonlinear Phenomena \textbf{160}(3), 222--254 (2001).
\newblock \doi{https://doi.org/10.1016/S0167-2789(01)00349-9}.
\newblock
  \urlprefix\url{https://www.sciencedirect.com/science/article/pii/S0167278901003499}

\bibitem{di2002bifurcations}
{di Bernardo}, M., Kowalczyk, P., Nordmark, A.: Bifurcations of dynamical
  systems with sliding: derivation of normal-form mappings.
\newblock Physica D: Nonlinear Phenomena \textbf{170}(3), 175--205 (2002).
\newblock \doi{https://doi.org/10.1016/S0167-2789(02)00547-X}.
\newblock
  \urlprefix\url{https://www.sciencedirect.com/science/article/pii/S016727890200547X}

\bibitem{feigin1978structure}
Feigin, M.: On the structure of c-bifurcation boundaries of
  piecewise-continuous systems: Pmm vol. 42, no. 5, 1978, pp. 820–829.
\newblock Journal of Applied Mathematics and Mechanics \textbf{42}(5), 885--895
  (1978).
\newblock \doi{https://doi.org/10.1016/0021-8928(78)90035-7}.
\newblock
  \urlprefix\url{https://www.sciencedirect.com/science/article/pii/0021892878900357}

\bibitem{george2016experimental}
George, C., Virgin, L.N., Witelski, T.: Experimental study of regular and
  chaotic transients in a non-smooth system.
\newblock International Journal of Non-Linear Mechanics \textbf{81}, 55--64
  (2016).
\newblock \doi{https://doi.org/10.1016/j.ijnonlinmec.2015.12.006}.
\newblock
  \urlprefix\url{https://www.sciencedirect.com/science/article/pii/S0020746215002413}

\bibitem{ing2006dynamics}
Ing, J., Pavlovskaia, E., Wiercigroch, M.: Dynamics of a nearly symmetrical
  piecewise linear oscillator close to grazing incidence: Modelling and
  experimental verification.
\newblock Nonlinear Dynamics \textbf{46}(3), 225--238 (2006).
\newblock \doi{10.1007/s11071-006-9045-9}.
\newblock \urlprefix\url{https://doi.org/10.1007/s11071-006-9045-9}

\bibitem{ing2007experimental}
Ing, J., Pavlovskaia, E., Wiercigroch, M., Banerjee, S.: Experimental study of
  impact oscillator with one-sided elastic constraint.
\newblock Philosophical Transactions of the Royal Society A: Mathematical,
  Physical and Engineering Sciences \textbf{366}(1866), 679--705 (2007).
\newblock \doi{https://doi.org/10.1098/rsta.2007.2122}.
\newblock
  \urlprefix\url{https://royalsocietypublishing.org/doi/10.1098/rsta.2007.2122}

\bibitem{ivanov1993stabilization}
Ivanov, A.: Stabilization of an impact oscillator near grazing incidence owing
  to resonance.
\newblock Journal of Sound and Vibration \textbf{162}(3), 562--565 (1993).
\newblock \doi{https://doi.org/10.1006/jsvi.1993.1142}.
\newblock
  \urlprefix\url{https://www.sciencedirect.com/science/article/pii/S0022460X83711429}

\bibitem{kundu2012singularities}
Kundu, S., Banerjee, S., Ing, J., Pavlovskaia, E., Wiercigroch, M.:
  Singularities in soft-impacting systems.
\newblock Physica D: Nonlinear Phenomena \textbf{241}(5), 553--565 (2012).
\newblock \doi{https://doi.org/10.1016/j.physd.2011.11.014}.
\newblock
  \urlprefix\url{https://www.sciencedirect.com/science/article/pii/S0167278911003344}

\bibitem{lenci1998procedure}
Lenci, S., Rega, G.: A procedure for reducing the chaotic response region in an
  impact mechanical system.
\newblock Nonlinear Dynamics \textbf{15}(4), 391--409 (1998).
\newblock \doi{10.1023/A:1008209513877}.
\newblock \urlprefix\url{https://doi.org/10.1023/A:1008209513877}

\bibitem{ma2006border}
Ma, Y., Agarwal, M., Banerjee, S.: Border collision bifurcations in a soft
  impact system.
\newblock Physics Letters A \textbf{354}(4), 281--287 (2006).
\newblock \doi{https://doi.org/10.1016/j.physleta.2006.01.025}.
\newblock
  \urlprefix\url{https://www.sciencedirect.com/science/article/pii/S0375960106000569}

\bibitem{ma2008nature}
Ma, Y., Ing, J., Banerjee, S., Wiercigroch, M., Pavlovskaia, E.: The nature of
  the normal form map for soft impacting systems.
\newblock International Journal of Non-Linear Mechanics \textbf{43}(6),
  504--513 (2008).
\newblock \doi{https://doi.org/10.1016/j.ijnonlinmec.2008.04.001}.
\newblock
  \urlprefix\url{https://www.sciencedirect.com/science/article/pii/S0020746208000668}.
\newblock Non-linear Dynamics of Engineering Systems

\bibitem{nordmark1991non}
Nordmark, A.: Non-periodic motion caused by grazing incidence in an impact
  oscillator.
\newblock Journal of Sound and Vibration \textbf{145}(2), 279--297 (1991).
\newblock \doi{https://doi.org/10.1016/0022-460X(91)90592-8}.
\newblock
  \urlprefix\url{https://www.sciencedirect.com/science/article/pii/0022460X91905928}

\bibitem{pavlovskaia2004analytical}
Pavlovskaia, E., Wiercigroch, M.: Analytical drift reconstruction for
  visco-elastic impact oscillators operating in periodic and chaotic regimes.
\newblock Chaos, Solitons \& Fractals \textbf{19}(1), 151--161 (2004).
\newblock \doi{https://doi.org/10.1016/S0960-0779(03)00128-0}.
\newblock
  \urlprefix\url{https://www.sciencedirect.com/science/article/pii/S0960077903001280}

\bibitem{pavlovskaia2004two}
Pavlovskaia, E., Wiercigroch, M., Grebogi, C.: Two-dimensional map for impact
  oscillator with drift.
\newblock Phys. Rev. E \textbf{70}, 036201 (2004).
\newblock \doi{10.1103/PhysRevE.70.036201}.
\newblock \urlprefix\url{https://link.aps.org/doi/10.1103/PhysRevE.70.036201}

\bibitem{peterka1992transition}
Peterka, F., Vacík, J.: Transition to chaotic motion in mechanical systems
  with impacts.
\newblock Journal of Sound and Vibration \textbf{154}(1), 95--115 (1992).
\newblock \doi{https://doi.org/10.1016/0022-460X(92)90406-N}.
\newblock
  \urlprefix\url{https://www.sciencedirect.com/science/article/pii/0022460X9290406N}

\bibitem{seth2020electronic}
Seth, S., Banerjee, S.: Electronic circuit equivalent of a mechanical impacting
  system.
\newblock Nonlinear Dynamics \textbf{99}(4), 3113--3121 (2020).
\newblock \doi{10.1007/s11071-019-05457-w}.
\newblock \urlprefix\url{https://doi.org/10.1007/s11071-019-05457-w}

\bibitem{shaw1983periodically}
Shaw, S., Holmes, P.: A periodically forced piecewise linear oscillator.
\newblock Journal of Sound and Vibration \textbf{90}(1), 129--155 (1983).
\newblock \doi{https://doi.org/10.1016/0022-460X(83)90407-8}.
\newblock
  \urlprefix\url{https://www.sciencedirect.com/science/article/pii/0022460X83904078}

\bibitem{suda2016does}
Suda, N., Banerjee, S.: Why does narrow band chaos in impact oscillators
  disappear over a range of frequencies?
\newblock Nonlinear Dynamics \textbf{86}(3), 2017--2022 (2016).
\newblock \doi{10.1007/s11071-016-3011-y}.
\newblock \urlprefix\url{https://doi.org/10.1007/s11071-016-3011-y}

\bibitem{thota2006continuous}
Thota, P., Dankowicz, H.: Continuous and discontinuous grazing bifurcations in
  impacting oscillators.
\newblock Physica D: Nonlinear Phenomena \textbf{214}(2), 187--197 (2006).
\newblock \doi{https://doi.org/10.1016/j.physd.2006.01.006}.
\newblock
  \urlprefix\url{https://www.sciencedirect.com/science/article/pii/S0167278906000200}

\bibitem{whiston1987global}
Whiston, G.: Global dynamics of a vibro-impacting linear oscillator.
\newblock Journal of Sound and Vibration \textbf{118}(3), 395--424 (1987).
\newblock \doi{https://doi.org/10.1016/0022-460X(87)90361-0}.
\newblock
  \urlprefix\url{https://www.sciencedirect.com/science/article/pii/0022460X87903610}

\bibitem{witkowski2019modelling}
Witkowski, K., Kudra, G., Wasilewski, G., Awrejcewicz, J.: Modelling and
  experimental validation of 1-degree-of-freedom impacting oscillator.
\newblock Proceedings of the Institution of Mechanical Engineers, Part I:
  Journal of Systems and Control Engineering \textbf{233}(4), 418--430 (2019).
\newblock \doi{10.1177/0959651818803165}.
\newblock \urlprefix\url{https://doi.org/10.1177/0959651818803165}

\end{thebibliography}
\end{document}